\documentclass[12pt,epsf,epsfig,psfig]{article}
\usepackage{graphicx}
\usepackage{epsfig}
\usepackage{cite}
\def\J{$J/\psi$}
\def\j{J/\psi}
\def\X{$\chi_c$}
\def\x{\chi}
\def\P{$\psi'$}

\def\U{$\Upsilon$}

\def\q{q{\bar q}}

\def\e{\epsilon}

\def\m{m_{\rm th}}

\def\be{\begin{equation}}
\def\ee{\end{equation}}

\def\lsim{\raise0.3ex\hbox{$<$\kern-0.75em\raise-1.1ex\hbox{$\sim$}}}
\def\gsim{\raise0.3ex\hbox{$>$\kern-0.75em\raise-1.1ex\hbox{$\sim$}}}

\def\NP{{ Nucl.\ Phys.\ }}
\def\PL{{ Phys.\ Lett.\ }}
\def\PR{{ Phys.\ Rev.\ }}

\def\PRL{{ Phys.\ Rev.\ Lett.\ }}

\def\ZP{{ Z.\ Phys.\ }}
\def\EP{{ Europ.\ Phys.\ J.\ }}

\begin{document}

15.03.09 \hfill BI-TP 2009/07 

\vskip 2cm

\centerline{\LARGE \bf The States of Matter in QCD}

\vskip 1.5cm 

\centerline{\large \bf Helmut Satz} 

\vskip 0.5cm

\centerline{Fakult\"at f\"ur Physik, Universit\"at Bielefeld, Germany}
 
\vskip2.5cm

\centerline{\bf Abstract:}

\bigskip

Quantum chromodynamics predicts that the interaction between its
fundamental constituents, quarks and gluons, can lead to different
states of strongly interacting matter, dependent on its temperature
and baryon density. We first survey the possible states of matter 
in QCD and discuss the transition from a color-confining hadronic 
phase to a plasma of deconfined colored quarks and gluons. Next, 
we summarize the results from non-perturbative studies of QCD at 
finite temperature and baryon density, and address the origin of 
deconfinement in the different regimes. Finally, we consider 
possible probes to test the basic features of bulk matter in QCD.

\newpage

\section{States of Strongly Interacting Matter}

\begin{figure}[htb]
\centerline{\psfig{file=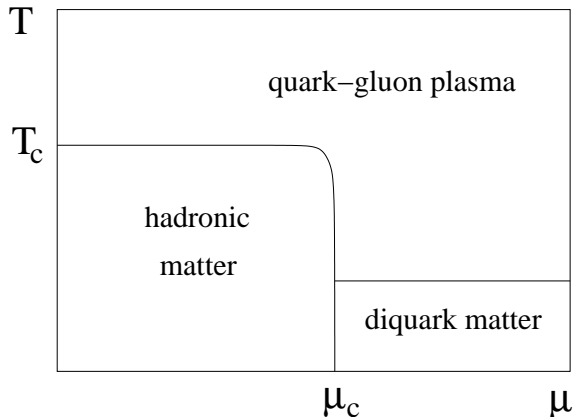,width=7.5cm}}
\caption{The phase diagram of QCD} 
\label{phase}
\end{figure}

What happens to strongly interacting matter in the limit of
high temperatures and densities? This question has fascinated
physicists ever since the discovery of the strong force and the 
multiple hadron production it leads to \cite{Heisenberg39,Fermi50,
Pomeranchuk51,Landau53,Hagedorn65}. Let us look at some of the 
features that have emerged over the years.
\begin{itemize}
\item{Hadrons have an intrinsic size, with a radius $r_h \simeq
1$ fm, and hence a hadron needs a space of volume $V_h \simeq (4\pi
/3)r_h^3$ in order to exist. This suggests a limiting
density $n_c = 1/V_h \simeq 2.4$ fm$^{-3}$ of hadronic matter 
\cite{Pomeranchuk51}. Beyond this point, hadrons overlap
more and more, so that eventually they cannot be identified any more.}
\vspace*{-0.2cm}
\item{Hadronic interactions provide abundant resonance production, and 
the resulting number $\rho(m)$ of hadron species increases
exponentially as function of the resonance mass $m$,
$\rho(m) \sim \exp(b~\!m)$. Such a form for $\rho(m)$ appeared first
in the statistical bootstrap model, based on self-similar resonance
formation or decay \cite{Hagedorn65}. It was then also obtained in the 
more dynamical dual resonance approach, which specifies the
scattering matrix through its pole structure \cite{DRM}. In hadron
thermodynamics, the exponential increase of the resonance degeneracy is found
to result in an upper limit for the temperature of hadronic matter, 
$T_c = 1/b \simeq 150-200$ MeV \cite{Hagedorn65}.}
\vspace*{-0.2cm}
\item{What happens beyond $T_c$? In QCD, the hadrons are dimensionful 
color-neutral bound states of the more basic pointlike colored quarks and
gluons. Hadronic matter, consisting of colorless constituents of hadronic
dimensions, can therefore turn at high temperatures and/or densities
into a quark-gluon plasma of pointlike colored quarks and gluons as 
constituents \cite{C-P}. This deconfinement transition leads to a 
color-conducting state and thus is the QCD counterpart of the 
insulator-conductor transition in atomic matter \cite{HS-Fort}.}
\vspace*{-0.2cm}
\item{A further transition phenomenon, also expected from the behavior of 
atomic matter, is a shift in the effective constituent mass. At $T=0$, 
in vacuum, quarks dress themselves with gluons to form the constituent 
quarks that make up hadrons. As a result, the bare quark mass $m_q \simeq 0$
is replaced by a constituent quark mass $M_q \sim 300$ MeV. In a
hot medium, this dressing melts and $M_q \to m_q$. Since the QCD
Lagrangian for $m_q=0$ is chirally symmetric, $M_q \not= 0$ implies
spontaneous chiral symmetry breaking. The melting $M_q \to 0$ thus
corresponds to chiral symmetry restoration. We shall see later on
that in QCD, as in atomic physics, the shift of the constituent
mass coincides with the onset of conductivity.}
\vspace*{-0.2cm} 
\item{So far, we have considered the ``heating'' of systems of low or 
vanishing baryon number density. The compression of baryonic matter
at low temperature could result in a third type of transition. This 
would set in if an attractive interaction between quarks in the deconfined 
baryon-rich phase results in the formation of colored bosonic diquark 
pairs, the counterpart of Cooper pairs in QCD. At sufficiently low
temperature, these diquarks can then condense to form a color
superconductor. Heating will dissociate the diquark pairs and turn
the color superconductor into a normal color conductor.}
\item{For a medium of quarks with color and flavor degrees of freedom, the 
diquark state can in fact consist of phases of different quantum 
number structures \cite{color-flavor}.
We also note that for increasing baryon density, the transition at low 
$T$ could lead to an intermediate ``quarkyonic'' state, in which baryons 
dissolve into quarks, but mesons remain as confined states 
\cite{quarky,hardcore}. In the present survey, we shall not pursue 
these interesting aspects any further.}
\end{itemize}
Using the baryochemical potential $\mu$ as a measure for the
baryon density of the system (i.e., for the total number of baryons
minus that of antibaryons, per unit volume), we then expect the phase diagram 
of QCD to have the general schematic form shown in Fig.\ \ref{phase}. 
Given QCD as the fundamental theory of strong interactions, we
can use the QCD Lagrangian as dynamics input to derive the resulting 
thermodynamics of strongly interacting matter. For vanishing 
baryochemical potential, $\mu=0$, this can be evaluated with the
help of the lattice regularisation, leading to finite temperature
lattice QCD. 

\section{From Hadrons to Quarks and Gluons}

Before turning to the study of strongly interacting matter in QCD,
we illustrate the 
transition from hadro\-nic matter to quark-gluon plasma by a very simple 
model. For an ideal gas of massless pions, the pressure as function of 
the temperature is given by the Stefan-Boltzmann form
\be
P_{\pi} = 3 {\pi^2 \over 90}~\! T^4 \label{2.3a}
\ee
where the factor 3 accounts for the three charge states of the pion.
The corresponding form for an ideal quark-gluon plasma with two
flavors and three colors is
\be
P_{qg} = \{ 2 \times 8 + {7\over 8}(3 \times 2 \times 2 \times 2) \}
{\pi^2 \over 90}~\! T^4 - B = 
37~\! {\pi^2 \over 90}~\! T^4 - B. \label{2.3b}
\ee
In Eq.\ (\ref{2.3b}), the first term in the curly brackets
accounts for the two spin and eight color degrees of freedom of the
gluons, the second for the three color, two flavor, two spin and two
particle-antiparticle degrees of freedom of the quarks, with 7/8 to
obtain the correct statistics. The bag pressure $B$ \cite{MIT}
takes into account the (non-perturbative) difference between the physical 
vacuum and the ground state for colored quarks and gluons \cite{Asakawa}.

\medskip

\begin{figure}[htb]
\mbox{
\hskip1.3cm
\epsfig{file=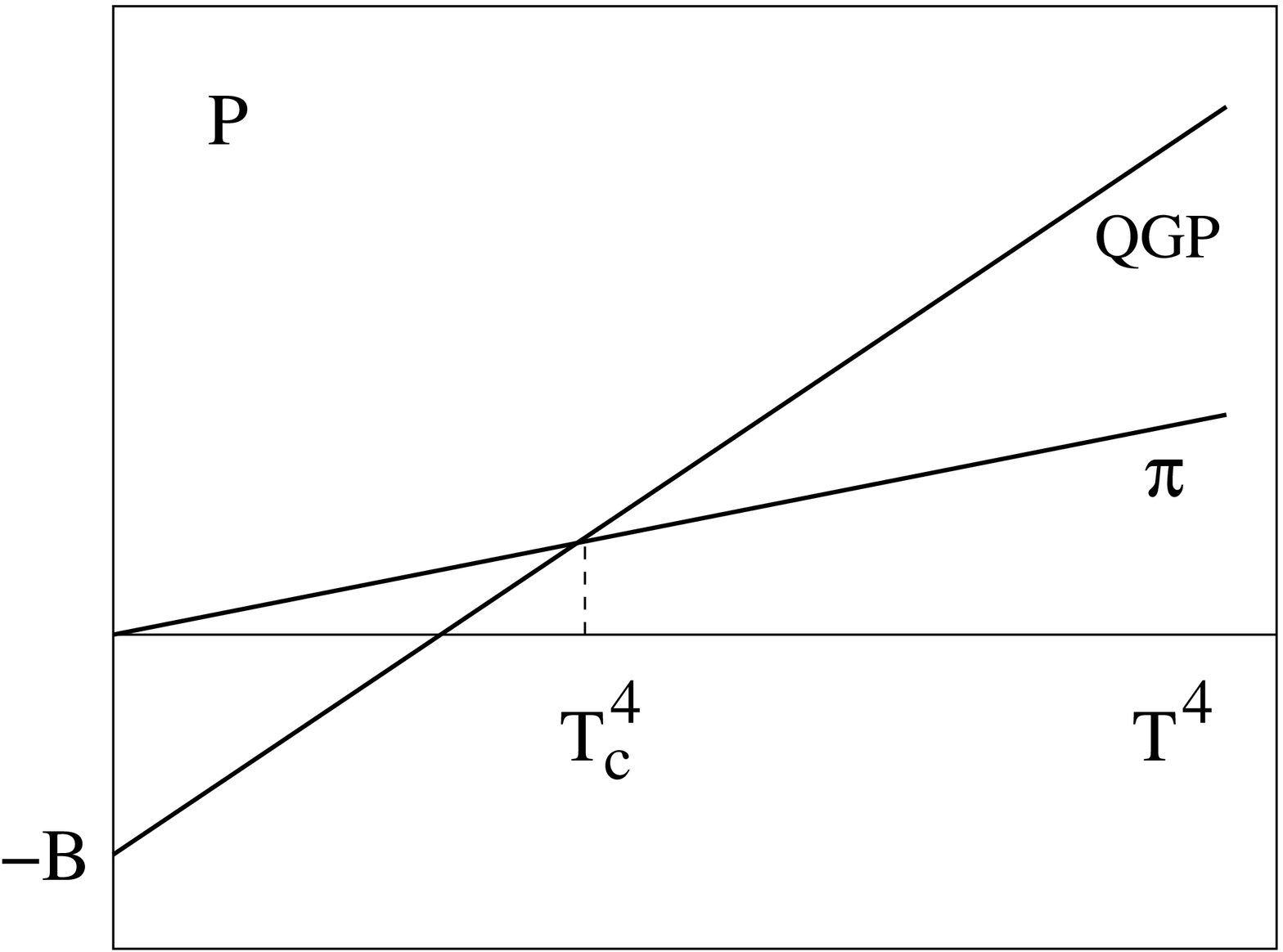,width=5.4cm}
\hskip2.5cm
\epsfig{file=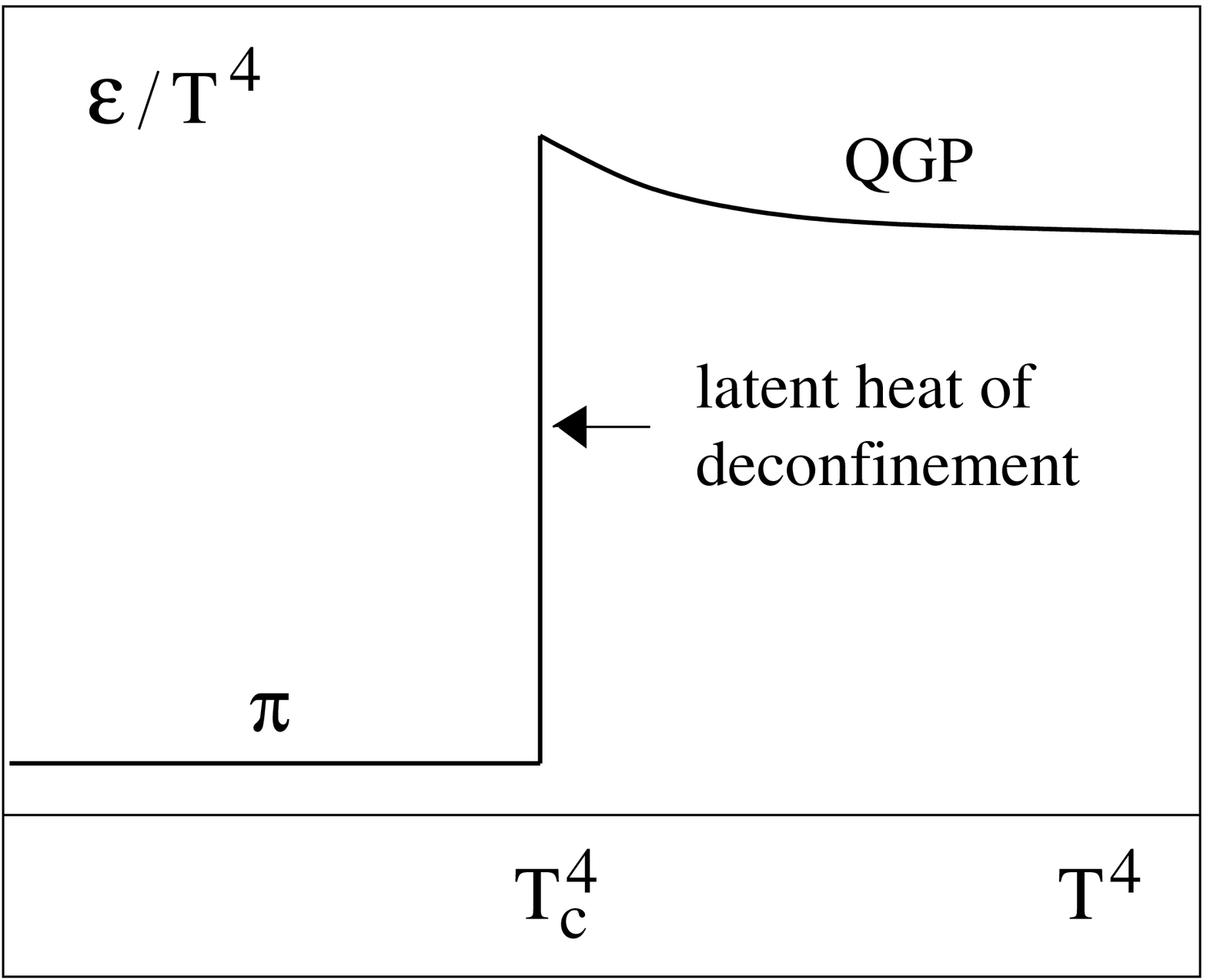,width=5cm}}
\vskip0.5cm
~~~~\hskip3.5cm (a) \hskip7cm (b) 
\caption{Pressure and energy density in a two-phase ideal gas
model.}
\label{2_3}
\end{figure}

Since in thermodynamics, a system chooses the state of lowest free
energy and hence highest pressure, we compare in Fig.\ \ref{2_3}$~\!$a the
temperature behavior of Eq's.\ (\ref{2.3a}) and (\ref{2.3b}). Our
simple model thus leads to a two-phase picture of strongly interacting
matter, with a hadronic phase up to
\be
T_c = \left( {45 \over 17 \pi^2} \right)^{1/4} B^{1/4}
 \simeq 0.72~B^{1/4} 
\label{Tc}
\ee
and a quark gluon plasma above this critical temperature. From hadron
spectroscopy, the bag pressure is given by $B^{1/4} \simeq 0.2$ GeV,
so that we obtain
\be
T_c \simeq 150~{\rm MeV} \label{2.3f}
\ee
as the deconfinement temperature. In the next section we shall find
this simple estimate to be remarkably close to the value obtained in
lattice QCD.

\bigskip

The energy densities of the two phases of our model are given by
\be
\e_{\pi} = {\pi^2 \over 10}~\! T^4 \label{2.3g}
\ee
and
\be
\e_{qg} = 37 {\pi^2 \over 30}~\! T^4 + B. \label{2.3h}
\ee
\vspace*{0.1cm}

By construction, the transition is first order, and the resulting
temperature dependence is shown in Fig.\ \ref{2_3}$~\!$b. At $T_c$,
the energy density increases abruptly by the latent heat of
deconfinement, $\Delta \e$. Using eq.\ (\ref{Tc}), its value is 
found to be
\be
\Delta \e = \e_{qg}(T_c) - \e_{\pi}(T_c) = 4~\!B,
\label{spec-heat}
\ee
so that it is determined completely by the bag pressure measuring the 
level difference between physical and colored vacua. 

\medskip

For an ideal gas of massless constituents, the trace $\e-3P$ of the 
energy-momentum tensor quite generally vanishes. Nevertheless, in our 
model of the ideal plasma of massless quarks and gluons, we have
for $T\geq T_c$
\be
\e - 3P = 4~\!B, 
\label{trace}
\ee

again specified by the bag pressure and not zero. This is related to 
the so-called trace anomaly and indicates the dynamical generation of a 
dimensional scale; we shall return to it in the next section, where we
will find that this scale is set by the vacuum expectation value of
the gluon condensate.

\section{Matter at Finite Temperature}

We now want to show that the conceptual considerations of the last
section indeed follow from strong interaction thermodynamics as based 
on QCD as the input dynamics. QCD is defined by the Lagrangian
\be
{\cal L}~=~-{1\over 4}F^a_{\mu\nu}F^{\mu\nu}_a~-~\sum_f{\bar\psi}^f
_\alpha(i \gamma^{\mu}\partial_{\mu} + m_f
-g \gamma^{\mu}A_{\mu})^{\alpha\beta}\psi^f_\beta
~,\label{2.4}
\ee
with
\be
F^a_{\mu\nu}~=~(\partial_{\mu}A^a_{\nu}-\partial_{\nu}A^a_{\mu}-
gf^a_{bc}A^b_{\mu}A^c_{\nu})~. \label{2.5}
\ee
Here $A^a_{\mu}$ denotes the gluon field of color $a$ ($a$=1,2,...,8)
and $\psi^f_{\alpha}$ the quark field of color $\alpha$
($\alpha$=1,2,3) and flavor $f$; the input (`bare') quark masses are
given by $m_f$, and $g$ is a dimensionless coupling.
With the dynamics thus determined, the corresponding
thermodynamics is obtained from the partition function, which is
most suitably expressed as a functional path integral,
\be
Z(T,V) = \int ~dA~d\psi~d{\bar\psi}~
\exp~\left(-\int_V d^3x \int_0^{1/T} d\tau~
{\cal L}(A,\psi,{\bar\psi})~\right), \label{2.6}
\ee
since this form involves directly the Lagrangian density defining the
theory. 
The spatial integration in the exponent of Eq.\ (\ref{2.6}) is
performed over the entire spatial volume $V$ of the system; in the
thermodynamic limit it becomes infinite. The time component $x_0$ is
``rotated" to become purely imaginary, $\tau = ix_0$, thus turning the
Minkowski manifold, on which the fields $A$ and $\psi$ are originally
defined, into a Euclidean space. The integration over $\tau$ in Eq.\
(\ref{2.6}) runs over a finite slice whose thickness is determined by
the temperature of the system; the vector (spinor) fields have to be
periodic (antiperiodic) at the boundary $\tau=0,\beta$.

\medskip

From $Z(T,V)$, all thermodynamical observables can be calculated in
the usual fashion. Thus
\be
\epsilon = \left( {T^2 \over V} \right)
\left({\partial \ln Z \over \partial T}\right)_V
\label{2.7}
\ee
gives the energy density, and
\be
P = T \left({\partial \ln Z\over \partial V}\right)_T
\label{2.8}
\ee
the pressure. For the study of critical behavior, long range
correlations and multi-particle interactions are of crucial importance;
hence perturbation theory cannot be used. The necessary
non-perturbative regularisation scheme is provided by the lattice
formulation of QCD \cite{Wilson}; it leads to a form which can be
evaluated numerically by computer simulation \cite{Creutz}.

\medskip

The calculational methods and techniques of finite temperature lattice QCD 
form a challenging subject on its own, which certainly surpasses the scope 
of this survey. We therefore restrict ourselves here to a summary of the 
main conceptual results obtained so far; for more details, we refer to 
the corresponding chapter of this handbook as well as to other excellent
books and reviews \cite{lattice}.

\medskip

\begin{figure}[htb]
\centerline{\psfig{file=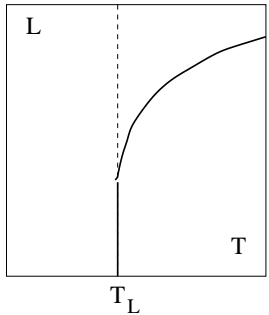,width=3.5cm}}
\caption{The temperature dependence of the Polyakov loop in pure
$SU(3)$ gauge theory.} 
\label{pol}
\end{figure}

The first variable considered in finite temperature lattice QCD 
is the deconfinement measure provided by the Polyakov loop \cite{Larry,Kuti}
\be
L(T) \sim \lim_{r \to \infty}~\exp\{-F(r)/T\} 
\label{polya}
\ee
where $F(r)$ is the free energy of a static quark-antiquark pair
separated by a distance $r$. In pure gauge theory, without light quarks,
$F(r) \sim \sigma r$, where $\sigma$ is the string tension; hence here 
$F(\infty)= \infty$ , so that $L=0$. In a deconfined medium, color
screening among the gluons leads to a melting of the string, which makes 
$F(r)$ finite at large $r$; hence now $L$ does not vanish. It thus becomes
an `order parameter' like the magnetisation in the Ising model: for
the temperature range $0 \leq T \leq T_L$, we have $L=0$ and hence 
confinement, while for $T_L < T$ we have $L>0$ and deconfinement;
see Fig.\ \ref{pol}.
The temperature $T_L$ at which $L$ becomes finite thus defines the
onset of deconfinement.

\medskip

In the large quark mass limit, QCD reduces to pure $SU(3)$ gauge theory, 
which is invariant under a global $Z_3$ symmetry. The Polyakov loop provides 
a measure of the state of the system under this symmetry: it vanishes for 
$Z_3$ symmetric states and becomes finite when $Z_3$ is spontaneously 
broken. Hence the critical behavior of $SU(3)$ gauge theory is in the 
same universality class as that of $Z_3$ spin theory (the 3-state Potts 
model): both are due to the spontaneous symmetry breaking of a global 
$Z_3$ symmetry, leading to a first order phase transition
\cite{Svetitsky}.

\medskip

For finite quark mass $m_q$, $F(r,T)$ remains finite for $r \to \infty$,
since the `string' between the two color charges `breaks' when the
corresponding potential energy becomes equal to the mass $M_h$ of the
lowest hadron; beyond this point, it becomes energetically more
favourable to produce an additional hadron. Hence now $L$ no longer
vanishes in the confined phase, but only becomes exponentially small
there,
\be
L(T) \sim \exp\{-M_h/T\}; 
\label{break}
\ee
here $M_h$ is a typical hadron mass, of the order of 0.5 to 1.0 GeV,
so that at $T_c \simeq 170$ MeV, $L \sim 10^{-2}$, rather than zero. 
Deconfinement is thus indeed much like the
insulator-conductor transition, for which the order parameter, the
conductivity $\sigma(T)$, also does not really vanish for $T>0$, but
with $\sigma(T) \sim \exp\{-\Delta E/T\}$ is only exponentially small,
since thermal ionisation (with ionisation energy $\Delta E$) produces
a small number of unbound electrons even in the insulator phase.

\medskip

\begin{figure}[htb]
\centerline{\psfig{file=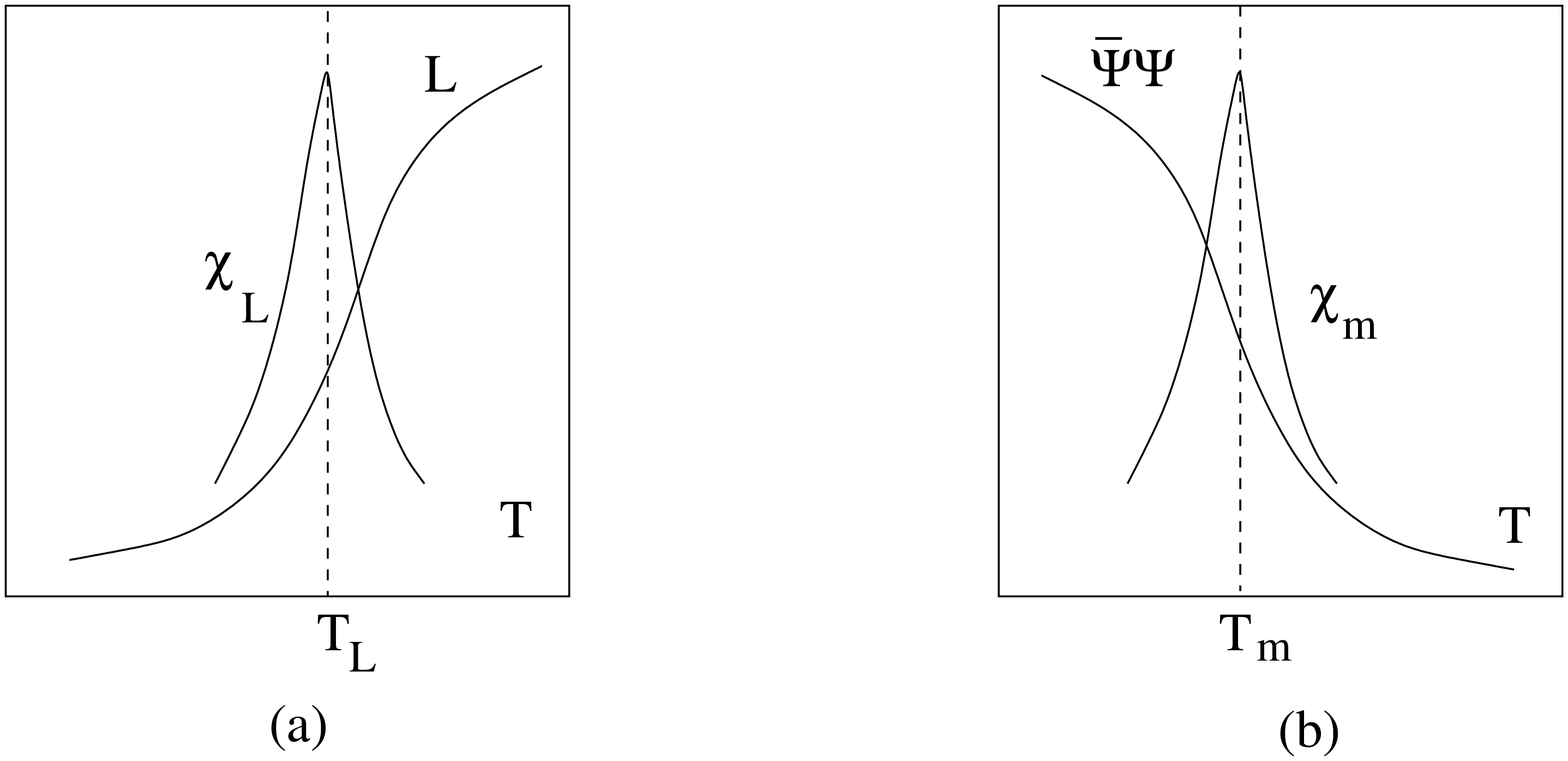,width=9.5cm}}
\caption{Schematic view of the temperature dependence of the 
Polyakov loop and of the chiral condensate, for $N_f=2$ and
small finite quark mass.}
\label{2_4}
\end{figure}

Fig.\ \ref{2_4}$~\!$a illustrates the schematically the behavior of 
$L(T)$ and of the corresponding susceptibility 
$\x_L(T) \sim \langle L^2 \rangle - \langle L \rangle^2$, 
as obtained in finite temperature lattice studies \cite{K&L,cheng75,
cheng77}, for the case of two flavors of light quarks.
We note that $L(T)$ undergoes the expected sudden increase from a small
confinement to a much larger deconfinement value. The sharp peak of
$\chi_L(T)$ defines quite well a transition temperature $T_L$, which
we shall shortly specify in physical units.


\medskip

\begin{figure}[htb]
\centerline{\psfig{file=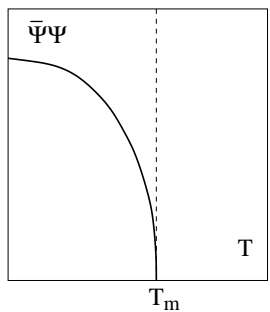,width=3.5cm}}
\caption{The temperature dependence of the 
chiral condensate in the limit $m_q=0$.} 
\label{chi}
\end{figure}

The next quantity to consider is the effective quark mass; it is
measured by the expectation value of the corresponding term in the
Lagrangian, $\langle {\bar \psi} \psi \rangle(T)$. In the
limit of vanishing current quark mass, the Lagrangian becomes chirally
symmetric and $\langle {\bar \psi} \psi \rangle(T)$ the corresponding
order parameter. In the confined phase, with effective constituent quark
masses $M_q \simeq 0.3$ GeV, this chiral symmetry is
spontaneously broken, while in the deconfined phase, at high enough
temperature, we expect its restoration. Hence now  
$\langle {\bar \psi} \psi \rangle(T)$ constitutes a genuine
order parameter, finite for $T< T_m$ and vanishing for $T\geq T_m$,
as shown in Fig.\ \ref{chi}.

\medskip

In the real world, with finite
pion and hence finite current quark mass, this symmetry is also only
approximate, since $\langle {\bar \psi} \psi \rangle (T)$ now never
vanishes at finite $T$.
The behavior of $\langle {\bar \psi} \psi \rangle(T)$ and of the
corresponding susceptibility $\chi_m \sim \partial \langle {\bar \psi}
\psi \rangle / \partial m_q$ are illustrated in Fig.\ \ref{2_4}$~\!$b,
again for two light quark flavors. We note here the
expected sudden drop of the effective quark mass and the associated
sharp peak in the susceptibility. The temperature $T_m$ at which this occurs
is generally found to coincide with the $T_L$ obtained through the 
deconfinement measure, leading to the conclusion that that at vanishing 
baryon number density, quark deconfinement and the shift from constituent 
to current quark mass define the same transition temperature $T_c$.  
However, one lattice group \cite{Fodor-Aoki} has recently found indications 
for two distinct transitions, with chiral symmetry being restored 
(at about 150 MeV) slightly before deconfinement occurs (at about 175 MeV). 
Such a behavior is very difficult to accommodate in most conventional 
confinement scenarios and hence must be investigated further.

\medskip

We thus obtain for $\mu_B=0$ a rather well defined phase structure,
consisting of a confined phase for $T < T_c$, with $L(T) \simeq 0$ and
$\langle {\bar \psi} \psi \rangle(T) \not= 0$, and a
deconfined phase for $T>T_c$ with $L(T)\not= 0$ and
$\langle {\bar \psi} \psi \rangle(T) \simeq 0$. The
underlying symmetries associated to the critical behavior at $T=T_c$,
the $Z_3$ symmetry of deconfinement and the chiral symmetry of the quark
mass shift, become exact in the limits $m_q \to \infty$ and $m_q \to
0$, respectively. In the real world, both symmetries are only
approximate; nevertheless, even for not too large finite quark masses,
both associated measures retain an almost critical behavior.

\medskip

\begin{figure}[htb]
\centerline{\psfig{file=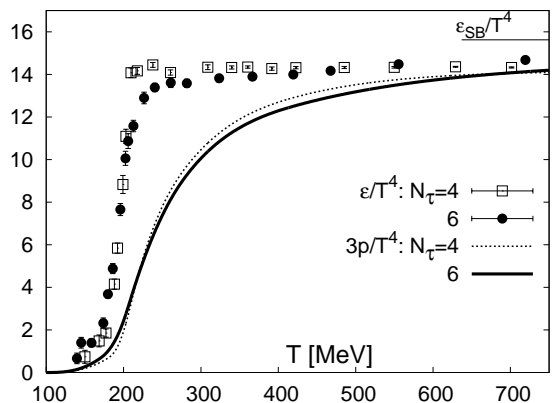,width=8.5cm}}
\caption{Energy density and pressure vs.\ temperature 
\cite{cheng77}}
\label{edens}
\end{figure}

Next we turn to the behavior of energy density $\e$ at deconfinement 
\cite{thermo}. In Fig.\ \ref{edens}, it is seen that for two light
and one heavy quark flavors, $\e/T^4$ changes quite
abruptly at the above critical temperature $T_c$, increasing from a low 
hadronic value to one slightly below that expected for an ideal gas of 
massless quarks and gluons \cite{thermo,Biele,cheng77}. 

\medskip

What is the value of the transition temperature? Since QCD
(in the limit of massless quarks) does not contain any dimensional
parameters, $T_c$ can only be obtained in physical units by expressing
it in terms of some other known observable which can also be calculated
on the lattice, such as the $\rho$-mass or the proton mass; more recently, 
the mass splitting of quarkonium states \cite{BBC} and 
the pion decay constant $f_\pi$ \cite{Fodor-Aoki} were used to gauge 
the lattice scale. Extrapolating the lattice results \cite{BBC} to the 
continuum limit and using charmonium splitting as scale led to 
$T_c\simeq 170 - 190$ MeV, while the results of \cite{Fodor-Aoki},
with two transitions, gave with $f_\pi$ as scale about 150 MeV for
chiral symmetry restoration and 175 MeV for deconfinement.

\medskip

\begin{figure}[htb]
\centerline{\psfig{file=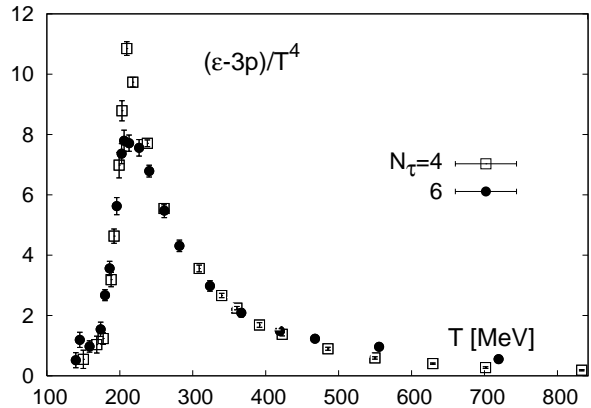,width=8.5cm}}
\caption{Interaction measure vs.\ tempe\-rature for two light and one
heavy quark flavors \cite{cheng77}} 
\label{inter}
\end{figure}

Related to the sudden increase of the energy density at deconfinement, there 
are two further points to note. In the region $T_c\!<\! T\! <\! 2~T_c$, 
there still remain strong interaction effects: the pressure does not 
show the same temperature dependence as $\e$, and hence the `interaction 
measure' $\Delta=(\e - 3P)/T^4$, shown in Fig.\ \ref{inter}, is sizeable 
and does not vanish, as it would for an ideal gas of massless constituents. 
In the simple model of the previous section, this effect arose due to the 
bag pressure, measuring the difference between the physical and the
``colored'' vacuum, and in actual QCD, one can also interpret it in such 
a fashion \cite{Asakawa}. More generally, it follows from the so-called
trace anomaly of QCD; let us consider this in a little more detail.

\medskip

As already noted, the QCD Lagrangian is scale-invariant: in the case of 
massless quarks and gluons it contains no dimensional scale. Any 
dimensional observable 
must therefore be measurable in terms of the temperature, and so the trace
of the energy-momentum tensor must be proportional to $T^4$: $\e-3P \sim T^4$.
This in turn implies that for $T\to 0$, it should vanish. However, the
vacuum expectation value of the gluon sector,
\be
G^2_0 \equiv \langle 0| F^a_{\mu\nu}F^{\mu\nu}_a |0 \rangle,
\label{g-condensate}
\ee
does not vanish there. Instead, it measures the sea of virtual gluons,
the gluon condensate, which defines the difference between the colored
system and the physical vacuum. This anomalous behavior was accounted
above through the bag pressure, and the numerical value of $G^2_0$ or
$B$ can only be determined (``gauged'') empirically, the theory itself being 
scale-invariant. In the MIT bag model \cite{MIT}, one obtains 
\be
R_N = \left( {3 \over 2 \pi B} \right)^{1/4} \! =
\left( {6 \over \pi G^2_0} \right)^{1/4}
\label{bagradius}
\ee
for the radius of a nucleon. For $R_N = 1$ fm, this leads to
$4 B = G_0^2 \simeq 1$ GeV/fm$^3$. There are various more refined
estimates for 
the gluon condensate at $T=0$, determined by different non-perturbative
hadronic inputs \cite{Shif}, giving a rather wide range of values, 
$G_0^2 \simeq 1 - 2$ GeV/fm$^3$ \cite{SVZ,Leut}. In any case,
to recover the correct vacuum physics, the trace of the energy momentum
tensor must be renormalized \cite{Leut}, giving
\be
\e - 3P = G_0^2 - G_T^2,
\label{T-condens}
\ee
where $G_T^2$ is the expectation value of the gluon condensate at
temperature $T$. In the confinement region $T \leq T_c$, the latter 
is found to remain close to $G_0^2$ \cite{Boyd,Adriano}, leading to 
$\e - 3P \simeq 0$,
until the temperature approaches the deconfinement point. There $G_T^2$ 
drops suddenly, i.e., the gluon condensate ``melts'' and as a consequence, 
the interaction measure shows a rapid rise, as seen in Fig.\ \ref{inter}.
We thus find once again that the temperature change of the gluon
condensate determines the specific heat of deconfinement: it appears 
that deconfinement corresponds to the melting of the gluon condensate. 

\medskip

The second point to note is that the thermodynamic observables remain 
about 10 \% below their Stefan-Boltzmann values (marked ``SB'' in Fig.\ 
\ref{edens}) even at very high temperatures, where the interaction measure
becomes very small. Such deviations from ideal gas 
behavior can be modelled in terms of effective `thermal' 
masses $m_{\rm th}$ of quarks and gluons, with $\m \simeq g(T)~T$
\cite{Golo,Patkos}. Maintaining the next-to-leading order term
in mass in the Stefan-Boltzmann form gives for the pressure
\be
P = c~\!T^4 \left[1 - a\left({\m\over T}\right)^2\right] = 
c~T^4 [1 - a~g^2(T)]
\ee
and for the energy density,
\be
\e = 3~\!c~\!T^4 \left[1 - {a\over 3}\left({\m\over T}\right)^2 -\!
{2a~\!\m \over 3~\!T}\left({\partial \m\over \partial T}\right)\right]
=3~\!c~\!T^4 \left[1 - a~\!g^2(T)
- {2 a~\! \m \over 3} \left({ \partial g\over \partial T}\right)\right],
\label{pert-edens}
\ee
where $c$ and $a$ are color- and flavor-dependent positive constants.
Since $g^2(T) \sim 1/\ln T$, we then obtain
\be
\Delta= -2 c~\! a~\! m_{\rm th} \left( {\partial g\over \partial T} \right)=
- c~\! a~\! \left({\partial g^2 \over \partial \ln T}\right) \sim - g^4
\label{pert-delta}
\ee
for the interaction measure. The deviations of $P/T^4$ and $\e/T^4$ from 
the massless Stefan-Boltzmann form thus vanish as $g^2\sim (\log~T)^{-1}$, 
while the interaction measure decreases more rapidly, vanishing as
$g^4 \sim (\log~T)^{-2}$. From eq.\ (\ref{pert-delta}) we also see that it 
is the running in $T$ of the coupling which brings in and describes 
the interaction; for a coupling ``constant'', we would have $\Delta=0$.

\medskip

In summary, finite temperature lattice QCD at vanishing overall
baryon density shows
\begin{itemize}
\item{that there is a transition leading to color deconfinement,
coincident with chiral symmetry restoration, at the temperature
$T_c \simeq$ 0.15 - 0.20 GeV;}
\vspace*{-0.2cm}
\item{that this transition is accompanied by a sudden increase in
the energy density  (the ``latent heat of deconfinement") from a small
hadronic value to a much larger value, about 10 \% below
that of an ideal quark-gluon plasma.}
\end{itemize}
In the two following sections, we want to address in more detail the
nature of the critical behavior encountered at the transition.

\section{The Order of the Transition}

We address here first the behavior of systems of vanishing overall baryon 
density ($\mu=0$) and come to the situation for $\mu\not=0$ at the end. 
Consider the case of three quark species $u,~d,~s$.
\begin{itemize}
\item{In the limit $m_q \to \infty$ for all quark species, we recover 
pure $SU(3)$ gauge theory, with a deconfinement phase transition provided 
by spontaneous $Z_3$ breaking. It is first order, as is the case for the 
corresponding spin system, the 3-state Potts model \cite{Svetitsky}.}
\vspace*{-0.2cm}
\item{For $m_q \to 0$ for all quark masses, the Lagrangian becomes 
chirally symmetric, so that we have a phase transition 
corresponding to chiral symmetry restoration. In the case of three
massless quarks, the transition is also of first order \cite{Pis-Wil}.}
\item{For $m_{u,d}=0$ and $m_s>m_s^{\rm tri}$, the transition is
of second order and presumably in the $O(4)$ universality class 
\cite{Pis-Wil}.
The second order limits of the first order regions appear to be in
the Ising ($Z_2$) universality class \cite{Ising}. 
At the tricritical point
$m_s^{\rm tri}$, the two different continuous transitions meet with
the first order transition \cite{tricrit}.}

\vspace*{0.2cm}

\begin{figure}[htb]
\centerline{\psfig{file=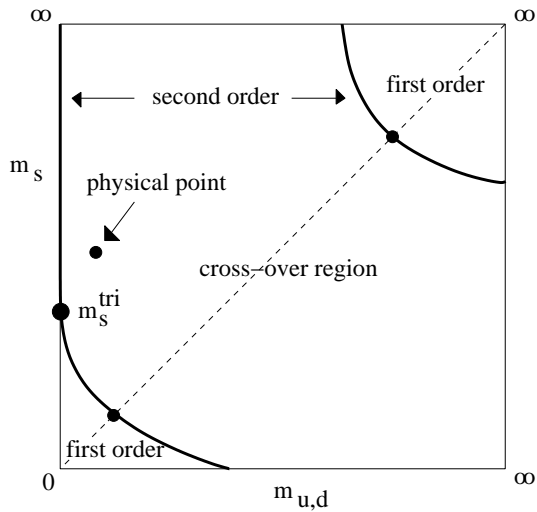,width=7cm}}
\caption{The form of thermal critical behavior in QCD}
\label{nature}
\end{figure}

\item{For $0 < m_q < \infty$, there is neither spontaneous $Z_3$
breaking nor chiral symmetry restoration. Hence in general, there is
no singular behavior, apart from the transient disappearence of the
first order discontinuities on a line of second order transitions.
Beyond this, there is no genuine phase transition, but
only a ``rapid cross-over'' from confinement
to deconfinement. The overall behavior is summarized in Fig.\ \ref{nature}.} 
\vspace*{-0.2cm}
\item{As already implicitely noted above, both ``order parameters''
$L(T)$ and $\chi(T)$ nevertheless
show a sharp temperature variation for all values
of $m_q$, so that it is in fact possible to define quite well a 
common cross-over point $T_c$.}
\vspace*{-0.2cm}
\item{The nature of the transition thus depends quite sensitively
on the number of flavors $N_f$ and the quark mass values: it can
be a genuine phase transition (first order or continuous), or just 
a rapid cross-over. The case realized in nature, the ``physical point'', 
corresponds to small $u,~d$ masses and a larger $s$-quark mass. 
It is fairly certain today that this point falls into the cross-over region.}
\vspace*{-0.2cm}
\end{itemize}

\begin{figure}[htb]
\centerline{\psfig{file=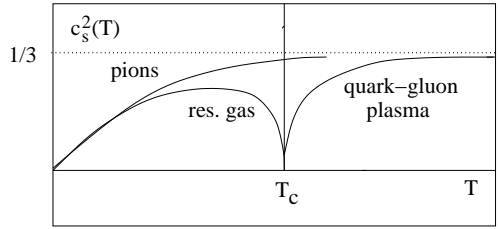,width=6.5cm}}
\caption{The speed of sound in QGP matter, compared to the
behavior of an ideal pion gas}
\label{sound-speed}
\end{figure}

Before turning to the behavior at finite baryon density, we want to
consider a particular consequence of the transition order.
The standard way to determine the order is a study
of the temperature dependence of the relevant order parameter; alternatively,
one may check the behavior of the energy density in the critical region.
Another, particularly instructive test is the speed of sound in the
interacting medium. It is defined as
\be
c_s^2(T) = \left({\partial P \over \partial \e}\right)_V =
\left({\partial P \over \partial T}\right)_V / 
\left({\partial \e \over \partial T}\right)_V
\label{s-speed}
\ee
and measures the relative change of the pressure compared to that of the
energy density. For an ideal gas of massless constituents, $c_s^2=1/3$.
In the confined state, we expect the system to behave 
like an ideal resonance gas, which for an exponentially increasing
resonance mass spectrum $\rho(m) \sim \exp\{b~\!m\}$ leads to critical
behavior at $T=T_c=1/b$. Very near $T_c$, any further energy input goes
into the production of more and heavier resonances, not into kinetic
energy, and hence $c_s^2$ has a pronounced minimum or vanishes
at $T_c$. In lattice
studies \cite{S-Redlich}, there are indications of such behavior, with
$c_s^2$ dropping as $T \to T_c$ both from below and from above $T_c$.
In the deconfined state, the decrease is physically less well understood.
The pattern is illustrated in Fig.\ \ref{sound-speed}.

\medskip

\medskip

\begin{figure}[h]
\hspace*{0.5cm}
\begin{minipage}[t]{6cm}
\epsfig{file=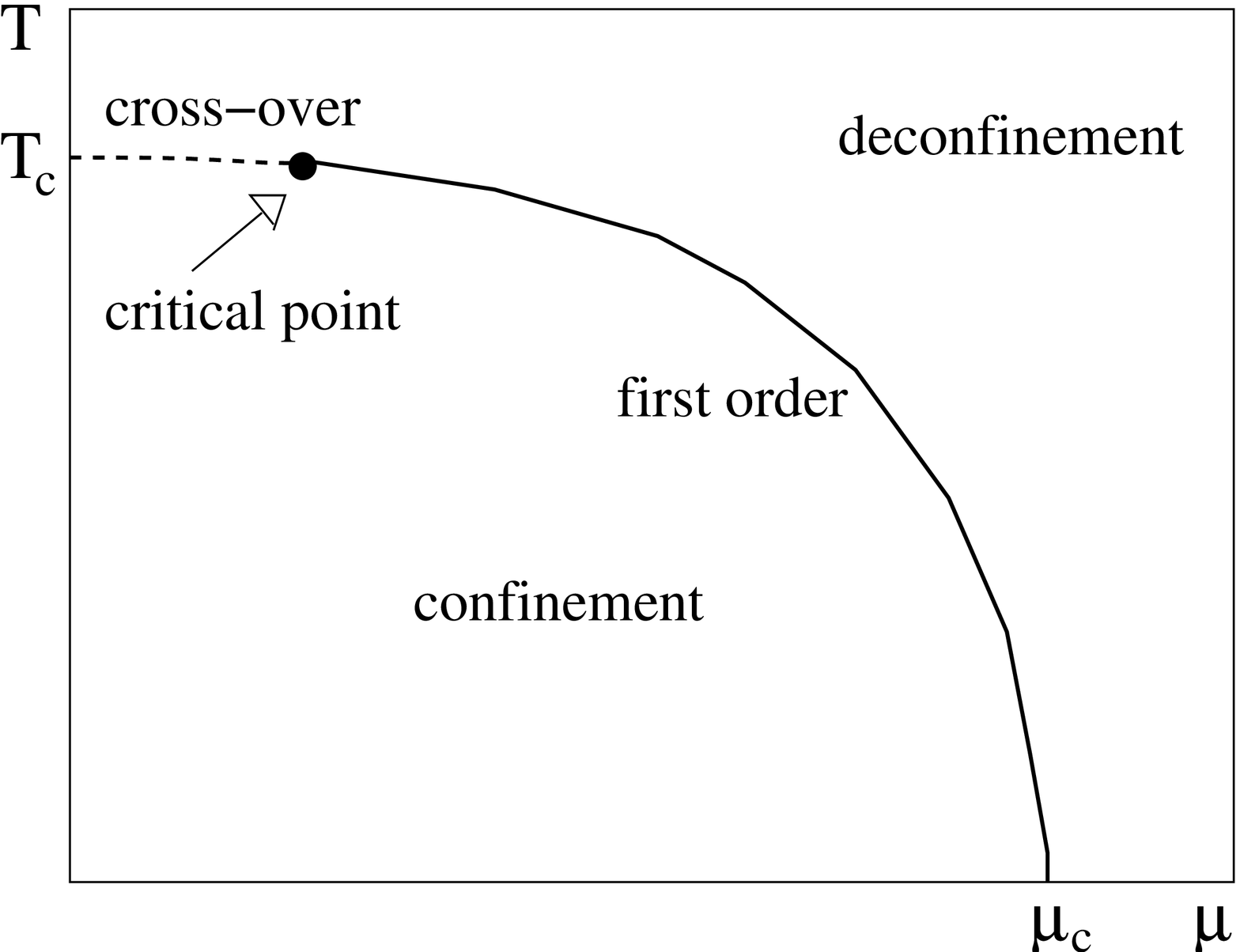,width=6.0cm,height=4.7cm}
\caption{Phase structure in terms of the baryon density}
\label{phase-d}
\end{minipage}
\hspace{2.9cm}
\begin{minipage}[t]{6.5cm}
\vspace{-5.0cm}
\hspace*{-1.0cm}
\epsfig{file=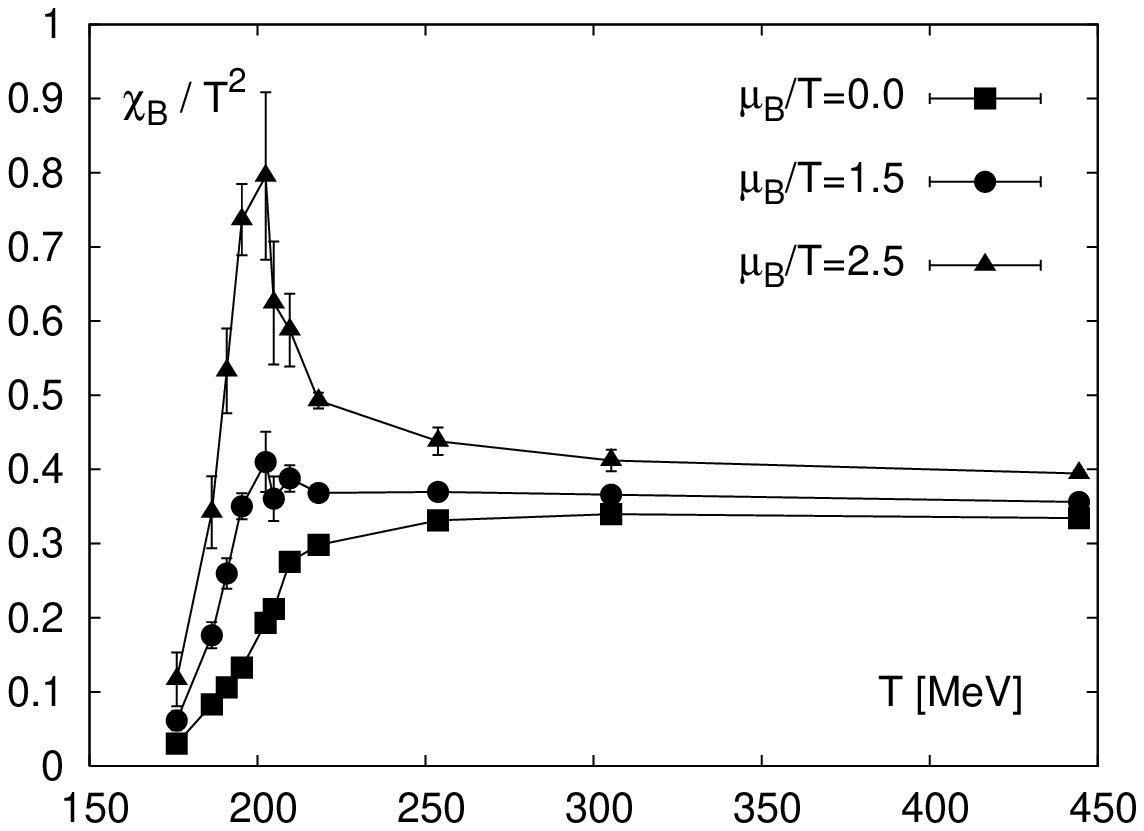,width=7.2cm,height=5.1cm}
\vskip-0.1cm
\caption{Baryon number sus-~~~~
ceptibility $\chi_q$ vs.\ temperature 
\cite{Swansea}} 
\label{fluc}
\end{minipage}
\end{figure}

Finally we want to consider the general phase diagram, allowing a
non-vanishing baryon density ($\mu\not=0$), assuming the number of 
baryons exceeds that of antibaryons. Here diverse general arguments
\cite{stepha} suggest for two light and one heavy quark flavors 
a phase diagram of the form  shown in Fig.\ \ref{phase-d}. 
It shows non-singular behavior in a region between 
$0 \leq \mu < \mu_c$, a critical point at $\mu_c$, and beyond this 
a first order transition. Unfortunately, for $\mu\not=0$ the conventional 
computer algorithms of lattice QCD break down, and hence new calculation 
methods have to be developed. First such attempts (reweighting \cite{Fodor}
or power series \cite{Swansea,GG}) are in accord with expected pattern;
thus the convergence radius of the power series expansion does seem
to be bounded. Further recent lattice calculation provide additional support;
as shown in Fig.\ \ref{fluc}, the baryon density fluctuations appear to
diverge for some critical value of the baryochemical potential 
\cite{Swansea,GG}. On the other hand, analytic continuation methods 
\cite{Forcrand,Lombardo} leave open the existence of any critical 
behavior at finite $\mu$.

\medskip

We conclude then that the critical behavior for strongly interacting matter 
at low or vanishing baryon density, describing the onset of confinement in 
the early universe as well as in high energy nuclear collisions, occurs 
in the rather enigmatic form of a ``rapid cross-over''. There is no 
thermal singularity and hence, in a strict sense, there are neither 
distinct states of matter nor phase transitions between them. So what 
does the often mentioned experimental search for a ``new state of matter'' 
really mean? How can a new state appear without a phase transition?
Is there a more general way to define and distinguish different states of bulk
media? After all, in statistical QCD one does find that thermodynamic 
observables -- energy and entropy densities, pressure, as well as 
the ``order parameters'' $L(T)$ and $\chi(T)$ -- continue to change rapidly 
and thus define a rather clear transition line in the entire cross-over 
region. Why is this so, what is the mechanism which can cause such a 
transition?
 
\section{The Origin of the Transition}

In the present section, we want consider a speculative answer to this
rather fundamental question \cite{hs-perco}, starting again with the
case of vanishing baryon density. The traditional phase 
transitions, such as the freezing of water or the magnetization of iron,
are due to symmetry breaking and the resulting singularities of the
partition function. But there are other ``transitions'', such as
making pudding or boiling an egg, where one also has two clearly
different states, but no singularities in the partition function.
Such ``liquid-gel'' transitions are generally treated in terms of
cluster formation and percolation \cite{Stauffer}. They also correspond 
to critical behavior, but the quantities that diverge are geometric (cluster
size) and cannot be obtained from the partition function. 

\medskip

\begin{figure}[htb]
\centerline{\psfig{file=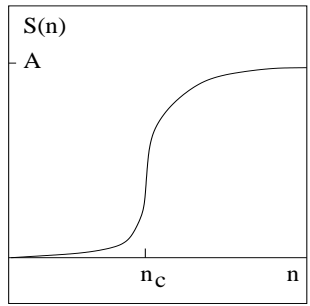,width=4cm,height=4cm}}
\caption{Cluster size $S(n)$ vs.\ density $n$}
\label{cluster}
\end{figure}

The simplest example of this phenomenon is provided by two-dimensional 
disk percolation. One distributes small disks of
area $a=\pi r^2$  randomly on a large surface $A=\pi R^2$, $R\gg r$, with
overlap allowed. With an increasing number of disks, clusters begin to
form. Given $N$ disks, the disk density is
$n=N/A$. Clearly, the average cluster $S(n)$ size will increase with $n$.
The striking feature is that it does so in a very sudden way (see
Fig.\ \ref{cluster}); as $n$ approaches some ``critical value'' $n_c$,
$S(n)$ suddenly becomes large enough to span the surface $A$. In fact, in the 
limit $N \to \infty$ and $A \to \infty$ at constant $n$, both
$S(n)$ and $dS(n)/dn$ diverge for $n \to n_c$: we have percolation
as a geometric form of critical behavior.

\medskip

The critical density for the onset of percolation has been determined
(numerically) for a variety of different systems. In two dimensions, disks 
percolate at $n_p\simeq 1.13/(\pi r^2)$, i.e., when we have a little more 
than one disk per unit area. Because of overlap, at this point only {68\%}
of space is covered by disks, 32\% remain empty. 
We therefore emphasize that $n_p$ is only the average
overall density for the onset of percolation. The density in the largest
and hence percolating cluster at this point must evidently be larger than
$n_p/0.68 \simeq 1.66/\pi r^2$; it is in fact found to be
$n_p^{\rm cl} \simeq 1.72/\pi r^2$ \cite{DFS}.

\medskip

In three dimensions, the corresponding problem is one of overlapping
spheres in a large volume. Here the critical density for the percolating 
spheres becomes $n_p \simeq 0.34/[(4\pi/3)r^3]$, with $r$ denoting the 
radius of the little spheres now taking the place of the small disks we 
had in two dimensions. At the critical point in three dimensions, however, 
only {29\%} of space is covered by overlapping spheres, while 71\% remains
empty, and here both spheres and empty space form infinite connected
networks. The density $n_p^{\rm cl}$ of the largest connected cluster at 
this point in overall density is thus much larger than $0.34/V_0$; in fact, 
it must exceed $1.17 /V_0$.

\medskip

Let us then consider hadrons of intrinsic size $V_h=(4\pi/3)r_h^3$, with 
$r_h \simeq 0.8$ fm. In three-dimensional space, the formation of a 
connected large-scale cluster first occurs at the overall average density
\be
n_c= {0.34 \over V_h} \simeq 0.16~{\rm fm}^{-3}.
\label{hadronmatter} 
\ee

This point specifies the onset of large-scale connected strongly 
interacting matter, 
in contrast to a gas of hadrons. However, as we saw, the density of the 
largest matter clusters is much higher than the average value given by 
eq.\ (\ref{hadronmatter}), and assuming all non-empty space to form
one cluster, we obtain $n_{\rm cl}\simeq 1.2/V_h \simeq 0.55~{\rm fm}^{-3}$ 
as (lower bound for the) critical density. Based on results for 
the two-dimensional case, we expect the threshold density $n_{\rm cl}$
to be about (1.5 - 2.0)/$V_h \simeq (0.7 - 0.9)~{\rm fm}^{-3}$.
If we assume that at this point, the cluster is of an ideal gas of all 
known hadrons and hadronic resonances, then we can calculate the temperature 
of the gas at the density $n_{\rm cl}$: $n_{\rm res}(T=T_c) = n_{\rm cl}$ 
implies $T_c \simeq 170 - 190$ MeV, which agrees quite well with the 
value of the deconfinement temperature found in lattice QCD for $\mu=0$. 
Cluster formation and percolation theory thus provide a possible tool 
to specify the deconfinement transition in strongly interacting matter.
 
\medskip

Such considerations may in fact well be of a more general nature than
the problem of states and transitions in strong interaction physics. 
The question of whether symmetry or connectivity (cluster formation)
determines the different states of many-body systems has intrigued
theorists in statistical physics for a long time \cite{F-K}. The
lesson learned from spin systems appears to be that cluster formation
and the associated critical behavior are the more general features, which 
under certain conditions can also lead to thermal criticality, i.e., 
singular behavior of the partition function. 

\medskip

Next we turn to the more general phase structure as function of
$T$ and $\mu$, as illustrated in Fig.\ \ref{phase-d}.
What conceptual aspects of hadronic interactions could lead to such 
behavior, and in particular, what features in hadronic dynamics result 
in the observed changes of the transition structure as function of 
baryon density?

\medskip

At low baryon density, the constituents of hadronic matter are mostly
mesons, and the dominant interaction is resonance formation; with increasing 
temperature, different resonance species of increasing mass are formed, 
leading to a gas of ever increasing degrees of freedom. They are all of 
a typical hadronic size (with a radius $R_h \simeq 1$ fm) and can 
overlap or interpenetrate each other. For $\mu \simeq 0$, the contribution 
of baryons/antibaryons and baryonic resonances is relatively small, but with 
increasing baryon density, they form an ever larger section of the species 
present in the matter, and beyond some baryon density, they become the 
dominant constituents. Finally, at vanishing temperature, the medium 
consists essentially of nucleons.

\medskip

\begin{figure}[htb]
\centerline{\psfig{file=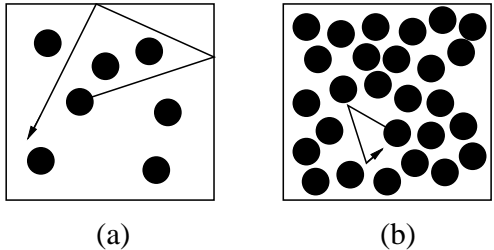,width=6.5cm}}
\caption{Hard sphere states: full mobility (a), ``jammed'' (b) \cite{KS}}
\label{hard}
\end{figure}

At high baryon density, the dominant interaction is non-resonant. 
Nuclear forces are short-range and strongly attractive at distances of 
about 1 fm; but for distances around 0.5 fm, they become strongly repulsive. 
The former is what makes nuclei, the latter (together with Coulomb and Fermi 
repulsion) prevents them from collapsing. The repulsion between a proton 
and a neutron shows the purely baryonic ``hard-core'' effect and is connected 
neither to Coulomb repulsion nor to Pauli blocking of nucleons. As a 
consequence, the volumes of nuclei grow linearly with the sum of its 
protons and neutrons. With increasing baryon density, the mobility of 
baryons in the medium becomes strongly restricted by the presence of other 
baryons (see Fig.\ \ref{hard}), leading to a ``jammed'' state in which 
each baryon can only move a small distance before being blocked by others 
\cite{KS}.

\medskip

\begin{figure}[htb]
\centerline{\psfig{file=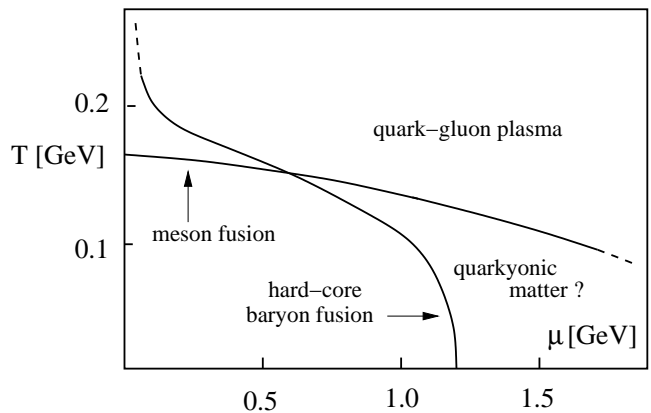,width=8.5cm}}
\caption{Fusion of mesonic bags vs.\ fusion of hard-core baryons 
\cite{hardcore}}
\label{hcperc}
\end{figure}

To addresss the situation of high baryon density, we again turn to
percolation theory, but now the constituents are hadrons containing
a repulsive hard core, which we take for simplicity to be half that
of the hadron. The percolation problem has been solved numerically
for such a case as well \cite{Kratky}. We thus have two percolation 
scenarios \cite{hardcore}: one for the ``bag fusion'' of
fully overlapping (or interpenetrating) 
mesonic spheres of radius $r_h\simeq 1$ fm, and one for baryons of 
the same radius, but having a hard core of radius $r_{hc}\simeq 0.5$ fm.  
In the $T-\mu$ plane, each percolation condition results in a transition
curve, as illustrated in Fig.\ \ref{hcperc}. As consequence, we have
for low $\mu$ a mesonic bag fusion transition to a quark-gluon plasma,
while for large $\mu$, the baryonic percolation transition is the first
to occur. It is thus quite conceivable that the competition between mesonic 
resonance clustering and the hard-core repulsion of baryons is at the origin
of the different transition patterns in the $T-\mu$ plane. Extending 
such a scenario even further, one may also consider the large $\mu$ region 
of the $T-\mu$ plane below the mesonic transition curve to become a 
further ``quarkyonic'' state of matter \cite{quarky}. 

\medskip

\section{Probing the States of Matter in QCD}

We thus find that at sufficiently high temperatures and/or densities,
strongly interacting matter will be in a new state, consisting of deconfined
quarks and gluons. Is there some way of studying this state experimentally?
The big bang theory for the creation and evolution of our universe implies
that in its early stages, it must have consisted of deconfined quark and
gluons. Neutron stars consist of very dense nuclear matter, and it is
conceivable that they have quark matter cores. Both these possible 
applications are interesting, yet they do not really allow a systematic
study. The rapid growth which 
the field has experienced in the past two decades was to a very large extent
stimulated by the idea that high energy nuclear collisions will
produce droplets of strongly interacting matter - droplets large enough
and long-lived enough to allow a study of the predictions which QCD makes
for macroscopic systems. Moreover, it is expected that the conditions provided 
in these interactions will suffice for quark plasma formation. Hence the
study of strongly interacting matter has today a multi-faceted experimental 
side; this, in turn, has stimulated much of the subsequent theoretical 
development. 

\medskip

The relevant experiments were initially denoted as ultra-relativistic 
nucleus-nucleus collisions; they are often, not quite correctly, also 
called heavy ion collisions (an ion fully stripped of its electrons is
a nucleus). The studies began at Brookhaven National 
Laboratory (BNL) near New York and at the European Center for Nuclear 
Research (CERN) near Geneva around 1986/87. The first collisions had 
light nuclei (oxygen, silicon, sulphur) hitting heavy targets (gold, 
uranium), since light ions could be dealt with using injectors already
existing at BNL and CERN. The successful analysis of these 
experiments provided the basis and motivation for the construction of
new injectors of truly heavy nuclei, gold at BNL and lead at CERN;
they came into operation in the middle 1990's. These early fixed target 
experiments were carried out at a center of mass energy of around 5 GeV 
per nucleon-nucleon collision at the BNL-AGS and around 20 GeV at the 
CERN-SPS. At the turn of millenium, the first dedicated nuclear accelerator,
the Relativistic Heavy Ion Collider RHIC, started taking data at BNL,
with a center of mass energy a factor ten higher, at 200 GeV per 
nucleon-nucleon collision. And coming soon now, the 
Large Hadron Collider LHC at CERN will bring the center-of-mass
energy for nuclear collisions up to 5500 GeV, or 5.5 TeV.

\medskip

The work of the different experimental groups working at these
facilities has provided an immense wealth of data, and there is
little doubt today that in such collisions comparatively large
systems of higher energy density are formed than have ever been
studied in the laboratory before. 
The detailed analyses of the results have also shown, however,
that a number of new features arise, features which go beyond
standard thermodynamics. Questions of non-equilibrium aspects, of
thermalization, evolution and expansion, cooling, flow and many more 
make a direct application of equilibrium strong interaction 
thermodynamics anything but straight-forward. Nevertheless, it
seems difficult to imagine that the more complex non-equilibrium
situation can be understood without having an understanding of
the simpler equilibrium case. Our aim here is therefore to 
address some of the more general problems which arise when one 
tries to study a bubble of strongly interacting matter which is
locally in equilibrium: what features allow us to determine the 
state of the matter inside such a bubble? We shall not address
here the really fundamental question of how a nuclear collision
can lead to equilibrated matter - this will be dealt with in
other chapters of this handbook.

\medskip

There are a number of methods we can use to analyse a sample of 
unknown strongly interacting matter:
\begin{itemize}
\item{hadron radiation,}
\item{electromagnetic radiation,}
\item{dissociation of a passing quarkonium beam,}
\item{energy loss of a passing hard jet.}
\end{itemize}
The first two are {\sl internal} probes, emitted by the thermal medium
itself, while the latter are in a sense {\sl external}: they are
expected to be formed by very early hard interactions,
before any thermal medium is established, and then test its
features by how their behavior is modified through its presence. 
All methods will be dealt with in detail in this handbook. Here we just 
want to summarize the essential ideas. 

\subsection{Hadron Radiation}

Consider a bubble of hot matter in a vacuum environment. Since the 
temperature of the bubble is
by assumption much higher than that of the environment, it will
radiate. Hadron radiation means that we study the emission of
hadrons consisting of light ($u,d,s$) quarks; their size is given by 
the typical hadronic scale of about 1 fm $\simeq$ 1/(200 MeV).
Since they cannot exist inside a deconfined medium, they are formed
at the transition surface between the hot matter and the physical vacuum.
The physics of this surface is independent of the interior - the
transition from deconfinement to confinement occurs at a temperature
$T \simeq\ 160- 190$ MeV, no matter how hot the medium initially was 
or still is in the interior of our volume. This is similar to having 
hot water vapor inside a glass container kept in a cool outside 
environment: at the surface, the vapor will condense into liquid, 
at a temperature of 100$^{\circ}$C - independent of the temperature
in the interior. As a result, studying soft hadron production in high 
energy collisions can provide us with information about the 
hadronization transition, but not about a hot QGP. 

\medskip

It should be noted here that the picture of a specific volume of hot 
matter, located in a vacuum and bounded by some surface, is just a
cartoon to illustrate the relevant phenomena. From the point of view of 
statistical physics, one should rather consider an infinite hot medium 
adiabatically cooling off. Hadronization will then occur locally 
everywhere, once the evolution of the medium reaches the critical 
point in temperature. 

\medskip

The state of the medium formed at hadronization is evidently an 
interacting system of hadrons. If we consider the medium to be
of low or even vanishing overall baryon density, the dominant
interaction is resonance formation and decay. In this case, the
interacting medium of basic hadrons (mainly pions, kaons, nucleons
and anti-nucleons) can be replaced by an ideal gas of all possible
resonances, both mesonic and baryonic \cite{B-U}. It was this 
concept that provided the basis for the statistical bootstrap
approach \cite{Hagedorn65} as well as of the dual-resonance model
\cite{DRM}. 

\medskip

Assuming then that the hadronization process is the formation 
of an ideal hadronic resonance gas, the relative abundances of the 
different species are determined \cite{thermal}.
The partition function of such a gas is for $\mu=0$ given by
\be 
\ln~Z(T,V) = \sum_{\rm hadrons~h} 
\ln~Z_h(T,V), 
\label{resgas} 
\ee
where 
\be 
\ln~Z_h(T,V) = d_h~ {V T \over 2 \pi^2}~m_h^2 K_2(m_h/T), 
\label{species} 
\ee
specifies the contribution of hadron or resonance species $h$, of
mass $m_h$ and (charge and spin) degeneracy $d_h$; we have here
assumed Boltzmann statistics. In an ideal
resonance gas of this type, the relative abundances of two species
$a$ and $b$ are predicted to be
\be
{N_a \over N_b} = {d_a m_a^2 K_2(m_a/T) \over d_b m_b^2 K_2(m_b/T)};
\label{abundances}
\ee
conservation laws have to be taken into account where applicable.
By studying the abundances of hadron species radiated by strongly
interacting matter, we thus obtain information about the hadronization 
temperature.

\medskip

One of the most striking observations in multihadron production 
in strong interaction physics is that the relative hadron 
abundances in all high energy collisions are correctly described
by this approach, from $e^+e^-$ annihilation to hadron-hadron and 
heavy ion interactions, and that they correspond to those of an 
ideal resonance gas at $T\simeq 170$ MeV \cite{Hagedorn65,thermal}. 
On the other hand, this raises the question of how 
``thermal'' hadronization 
actually occurs: in $e^+e^-$ annihilation one cannot really consider
the formation of strongly interacting ``matter'' as origin. Recent
studies have therefore related thermal hadron production more generally
to the existence of a color event horizon, allowing only tunnelling of
thermal signals to the outside world \cite{CKS}. This would make such 
production the QCD counterpart of Hawking-Unruh radiation from black 
holes \cite{H-U}.

\medskip

Hadron radiation, as we have pictured it here, is oversimplified from
the point of view of heavy ion interactions. In
the case of static thermal radiation, at the point of hadronization
all information about the earlier stages of the medium is lost, as we
had noted above. If, however, the early medium has a very high energy
density and can expand freely, i.e., is not constrained by the walls
of a container, then this expansion will lead to a global hydrodynamic 
flow \cite{Landau},
giving an additional overall boost in momentum to the produced hadrons:
they will experience a ``radial flow'' depending on the initial energy 
density. Moreover, if the initial conditions were 
not spherically symmetric, as is in fact the cases in peripheral heavy
ion collisions, the difference in pressure in different spatial
directions will lead to a further ``directed'' or ``elliptic'' flow.
Both forms of flow thus do depend on the initial conditions.
While the abundances of the species are not affected by such flow aspects,
the different momentum distributions are and hence studies of hadron spectra 
can, at least in principle, provide information about the earlier,
pre-hadronic stages. 

\subsection{Electromagnetic Radiation}

The hot medium also radiates electromagnetically, i.e., it
emits photons and dileptons ($e^+e^-$ or $\mu^+\mu^-$ pairs) \cite{e-m}. 
These are formed either by the interaction of quarks and/or gluons, or by
quark-antiquark annihilation. Since the photons and leptons interact
only electromagnetically, they will, once they are formed, leave the medium 
without any further modification. Hence their spectra provide information
about the state of the medium at the place or the time they were formed, 
and this can be in its deep interior or at very early stages of its
evolution. Photons and dileptons thus provide a possible probe of the 
hot QGP. The only problem is that they can be formed anywhere and at any
time, also at later evolution stages and as well as through interaction or
decay of the emitted hadrons. The task in making electromagnetic radiation
a viable tool is therefore the identification of the hot ``thermal'' 
radiation indeed emitted by the QGP. 

\medskip

\begin{figure}[htb]
\vspace*{-0mm}
\centerline{\psfig{file=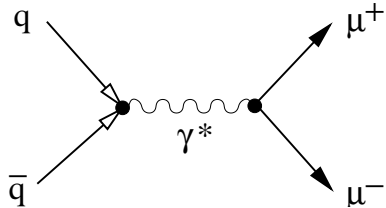,width=5cm}}
\caption{Dilepton production through $\q$ annihilation }
\label{anni}
\end{figure}

For the production of dilepton pair (for illustration, we consider
$\mu^+\mu^-$) by a thermal medium, the lowest
order process is quark-antiquark annihilation, as illustrated in Fig.\
\ref{anni}. To calculate the mass spectrum of the emitted dileptons,
the perturbative annihilation cross-section 
$\sigma(q \bar q \to \mu^+\mu^-)$ has to be convoluted by
thermal quark and antiquark momentum distributions 
\be
f(k_q/T) \sim \exp\{ -|k_q| /T\},
\ee
where $k_q$ is the three-momentum of the (massless) quark and $T$
the temperature of the medium. We thus
obtain
\be
{dN \over dM} \sim \int d^3 k_q f(k_q) d^3 k_{\bar q} f(k_{\bar q})
~\!\sigma(q \bar q \to \mu^+\mu^-),
\ee
where $M$ is the invariant mass of the dilepton. The convolution
leads to the schematic result
\be
 {dN \over dM} \sim \exp\{-M/T\},
\ee
so that a measurement of a thermal dilepton spectrum provides the
temperature of the medium. As already indicated, if the medium
undergoes an evolution (cooling), the observed dileptons originate
from all stages, so that a temperature measurement is not straight-forward
and will in general depend on the evolution pattern. In actual 
nuclear collision experiments, there is in addition competition from
non-thermal dileptons (from hard primary Drell-Yan production at
large $M$ and from hadronic decays at lower $M$). 

\medskip

\begin{figure}[htb]
\vspace*{-0mm}
\centerline{\psfig{file=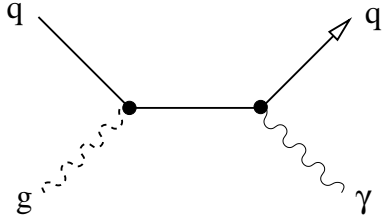,width=5cm}}
\caption{Photon production through gluonic Compton scattering}
\label{Compton}
\end{figure}

For photon production, the situation is similar. Here the dominant
process is a gluonic Compton effect, as illustrated in Fig. \ref{Compton}.
The rate is now given by a convolution of a thermal quark with thermal
gluon distribution, integrating over the perturbative Compton cross
section $\sigma(q g  \to q \gamma)$. The result is
\be
 {dN \over d\omega} \sim \exp\{-\omega/T\},
\ee
where $\omega = \sqrt p^2_g$ denotes the energy and $p$ the momentum
of the emitted gluon. Here again the two basic problems of electromagnetic
probes arise: the thermal photons originate from all evolution stages
and are in competition from non-thermal sources, both ``prompt'' hard
photons and hadronic decay products. 

\medskip

In either case the crucial signal to search for is a temperature-dependence
of the mass or momentum spectra. If an increase of the parametric 
temperature of the spectra with collision energy could be found, these
could indicate the production of media of increasing initial temperature.

\subsection{Quarkonium Dissocation}

The quark-gluon plasma consists by definition of deconfined and hence
colored gluons, quarks and anti-quarks. One of the essential features
of an electromagnetic plasma is Debye charge-screening, which reduces
the long-range Coulomb potential in vacuum to a much shorter range screened 
in-medium form,
\be
{e^2 \over r} \to {e^2 \over r}~ \exp\{-\mu~\! r\},
\label{D-screen}
\ee
where $\mu$ is the screening mass specifying the Debye or screening
radius $r_D=1/\mu$. In a plasma of color-charged constituents, one
expects a similar behavior, and this is indeed observed in lattice
studies \cite{color-screen}: in the QGP just above $T_c$, $\mu$ 
increases strongly, more than linearly, and hence
$r_D$ decreases correspondingly. Asymptotically, perturbation
theory suggests $\mu \simeq g(T) T$, with $g(T)$ for the strong
coupling running in temperature. The range of strong interactions
thus shows a striking in-medium decrease for increasing temperature.   

\medskip
 
Quarkonia are a special kind of hadrons, bound states of a heavy ($c$ or
$b$) quark and its antiquark. For the ground states \J~ and \U~the
binding energies are around 0.6 and 1.2 GeV, respectively, and thus
much larger than the typical hadronic scale $\Lambda \sim 0.2$ GeV;
as a consequence, they are also much smaller, with radii $r_Q$
of about 0.1 and 0.2 fm.  The fate of such states in a quark-gluon 
plasma therefore depends on the relative size of the color screening 
radius: if $r_D \gg r_Q$, the medium does not really affect the 
heavy quark binding. Once $r_D \ll r_Q$, however, the two heavy quarks
cannot ``see'' each other any more and hence the bound state will melt
\cite{MS}. It is therefore expected that quarkonia will survive in a 
quark-gluon plasma through some range of temperatures above $T_c$, and 
then melt once $T$ becomes large enough. Such behavior is
in fact confirmed by finite temperature lattice QCD 
studies of in-medium quarkonium behavior \cite{Psi-Lat}. 

\medskip

The higher excited quarkonium states are less tightly bound and hence 
larger, although their binding energies are in general still larger, 
their radii still smaller, than those of the usual light quark hadrons. 
Take the charmonium spectrum as example: the radius of the \J(1S) is 
about 0.2 fm, that of the \X(1P) about 0.3 fm, and that of the \P(2S) 
0.4 fm. Since melting sets in when the screening radius reaches the
binding radius, We expect that the different charmonium 
states have different ``melting temperatures'' in a quark-gluon plasma. 
Hence the spectral analysis of in-medium quarkonium dissociation should 
provide a QGP thermometer \cite{KMS,KSa}. 

\medskip

\begin{figure}[htb]
\vspace*{0.5cm}
{~~~~~~~~~~~~~~~~~~~~~~\psfig{file=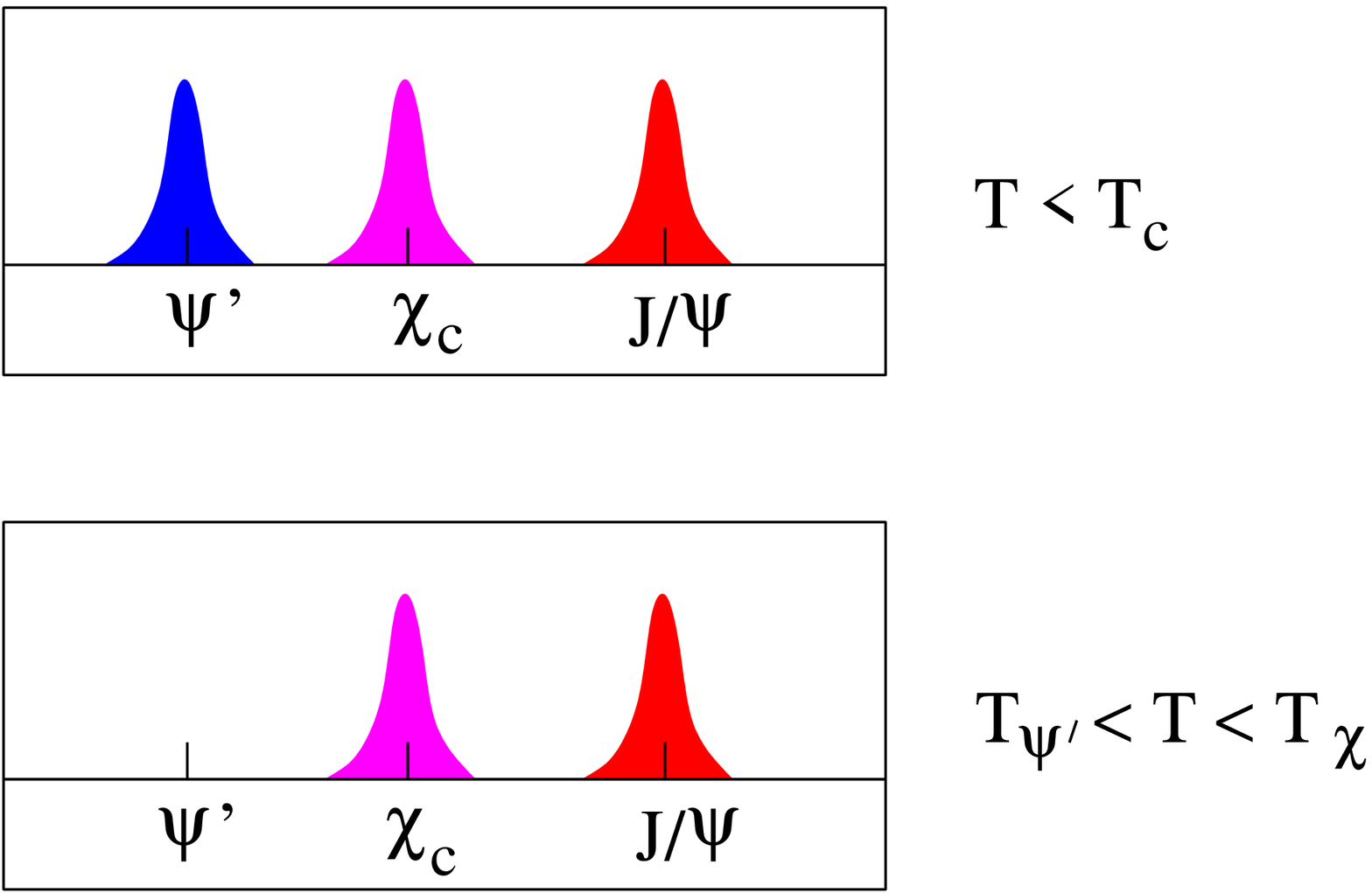,width=5.4cm}}
\hfill 
\vspace*{-3cm}
{\psfig{file=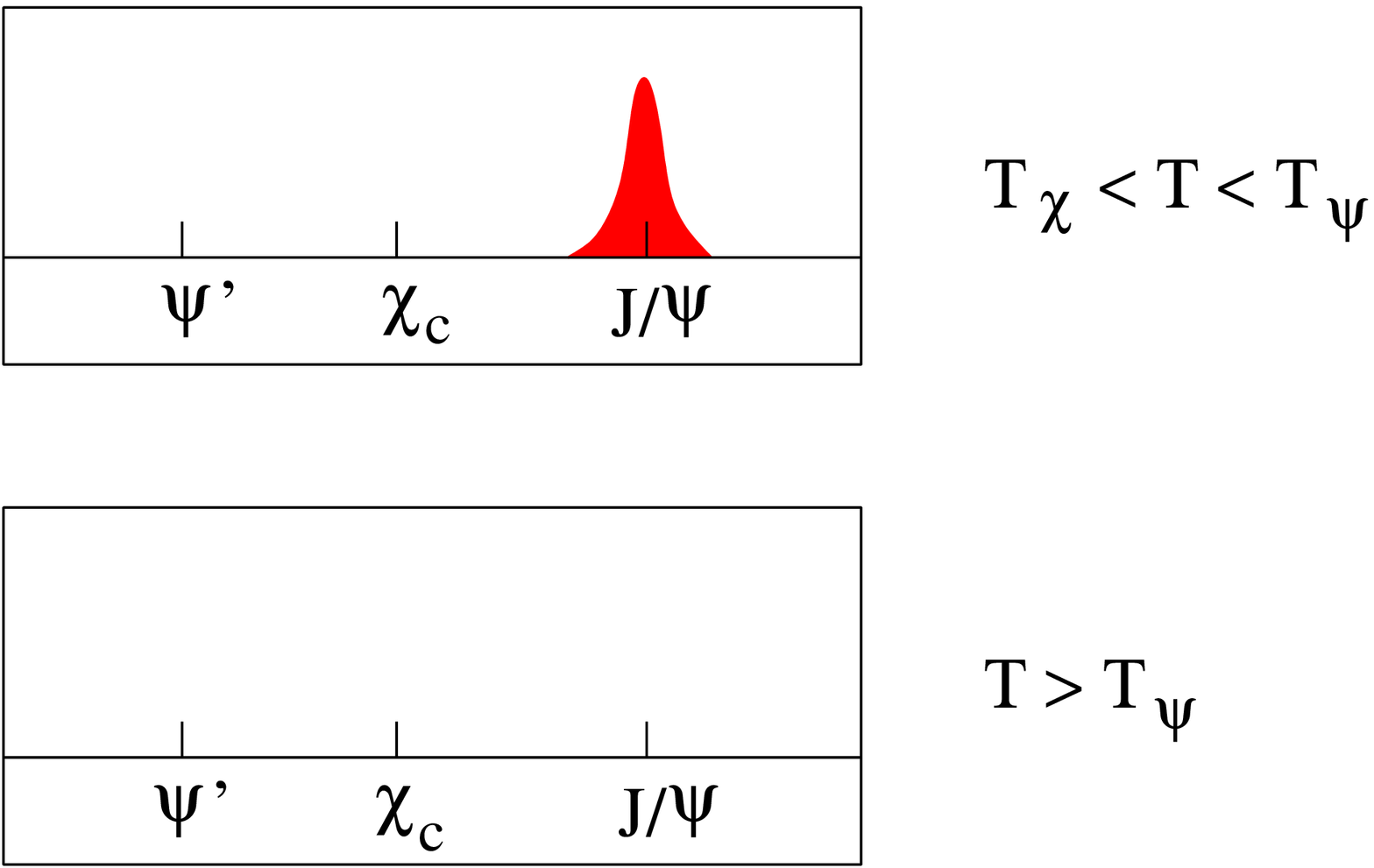,width=5.5cm}~~~~~~~~~}
\vspace*{3.5cm}
\caption{Charmonia as thermometer}
\label{temp}
\end{figure}

As probe, we then shoot beams of specific charmonia (\J,~\X,~\P)
into our medium sample and check which comes out on the other side.
If all three survive, we have an upper limit on the temperature, and by
checking at just what temperature the \P, the \X~and the \J~are
dissociated, we have a way of specifying the temperature of the 
medium \cite{KSa}, as illustrated in Fi.\ \ref{temp}.

\medskip

The dissociation of quarkonium states in a deconfined medium, as
compared to their survival in hadronic matter, can also be considered
on a more dynamical level, using the \J~as example. The \J~is a hadron 
with characteristic short-distance features; in particular, rather 
hard gluons are necessary to resolve or dissociate it, making such 
a dissociation accessible to perturbative calculations. \J~collisions 
with ordinary hadrons made up of the usual $u,d$ and $s$ quarks thus 
probe the local partonic structure of these `light' hadrons, not their 
global hadronic aspects, such as mass or size. It is for this reason 
that \J's can be used as a confinement/deconfinement probe.

\medskip

This can be illustrated by a simple example. Consider an ideal pion gas
as a confined medium. The momentum spectrum of pions has the Boltzmann
form $f(p) \sim \exp-(|p|/T)$, giving the pions an average momentum
$\langle |p| \rangle = 3~T$. With the pionic gluon distribution function
$xg(x) \sim (1-x)^3$, where $x=k/p$ denotes the fraction of the pion
momentum carried by a gluon, the average momenta of gluons confined to
pions becomes
\be
\langle |k| \rangle_{\rm conf}  \simeq 0.6~T. \label{5}
\ee
On the other hand, an ideal QGP as prototype of a deconfined medium
gives the gluons themselves the Boltzmann distribution $f(k) \sim
\exp-(|k|/T)$ and hence average momenta
\be
\langle |k| \rangle_{\rm deconf} = 3~T. \label{6}
\ee
Deconfinement thus results in a hardening of the gluon momentum
distribution. More generally speaking, the onset of deconfinement will
lead to parton distribution functions which are different from those
in vacuum, as determined by deep inelastic scattering experiments.
Since hard gluons are needed to resolve and dissociate \J's, one can use
\J's to probe the in-medium gluon hardness and hence the confinement
status of the medium.

\medskip

\begin{figure}[htb]
\vspace*{-0mm}
\centerline{\psfig{file=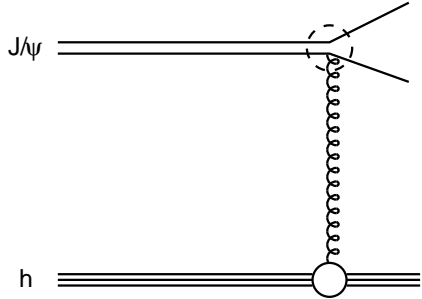,height=40mm}}
\caption{\J~dissociation by hadron interaction.}
\label{4_1}
\end{figure}

\begin{figure}[tbp]
\vspace*{-0mm}
\centerline{\psfig{file=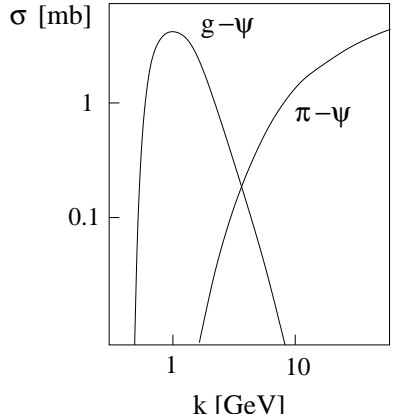,height=5.5cm}}
\caption{\J~dissociation by gluons and by pions; $k$ denotes the
momentum of the projectile incident on a stationary \J.}
\label{4_2}
\end{figure}

These qualitative considerations can be put on a solid theoretical basis
provided by short-distance QCD \cite{Peskin,Kaidalov,KS3}. In
Fig.\ \ref{4_1} we show the relevant diagram for the calculation
of the inelastic
\J-hadron cross section, as obtained in the operator product expansion
(essentially a multipole expansion for the charmonium quark-antiquark
system). The upper part of the figure shows \J~dissociation by gluon
interaction; the cross section for this process,
\be
\sigma_{g-\j} \sim (k-\Delta E_{\psi})^{3/2}  k^{-5}, \label{7}
\ee
constitutes the QCD analogue of the photo-effect. Convoluting the \J~
gluon-dissociation with the gluon distribution in the incident hadron,
$xg(x) \simeq 0.5(1-x)^{1+n}$, we obtain
\be
\sigma_{h-\j} \simeq \sigma_{\rm geom} (1 - \lambda_0/\lambda)^{n+3.5}
\label{8}
\ee
for the inelastic \J-hadron cross section, with $\lambda \simeq
(s-M_{\psi}^2)/M_{\psi}$ and $\lambda_0 \simeq (M_h + \Delta E_{\psi}$);
$s$ denotes the squared \J-hadron collision energy. In Eq.\ (\ref{8}),
$\sigma_{\rm geom} \simeq {\rm const}.\ r_{\psi}^2 \simeq 2 - 3$ mb is
the geometric cross section attained at high collision energies with the
mentioned gluon distribution. In the threshold region and for
relatively low collision energies, $\sigma_{h-\j}$ is very strongly
damped because of the suppression $(1-x)^{1+n}$ of hard gluons in
hadrons, which leads to the factor $(1 - \lambda_0/\lambda)^{n+3.5}$ in
Eq.\ (\ref{8}). In Fig.\ \ref{4_2}, we compare the cross sections
for \J~dissociation by gluons (``gluo-effect") and by pions ($n=2$), as
given by Eq's (\ref{7}) and (\ref{8}).
Gluon dissociation shows the typical photo-effect form, vanishing until
the gluon momentum $k$ passes the binding energy $\Delta E_{\psi}$;
it peaks just a little later and then vanishes again when
sufficiently hard gluons just pass through the much larger charmonium
bound states. In contrast, the \J-hadron cross section remains
negligibly small until rather high hadron momenta (3 - 4 GeV). In a
thermal medium, such momenta correspond to temperatures of more than one
GeV. Hence confined media in the temperature range of a few hundred MeV
are essentially transparent to \J's, while deconfined media of the
same temperatures very effectively dissociate them and thus are
\J-opaque.

\subsection{Jet Quenching}

Another possible probe is to shoot an energetic parton, quark or gluon, 
into our medium to be tested. How much energy it loses when it comes out
on the other side will tell us something about the density of the
medium \cite{Bj}. In particular, the density in a quark-gluon plasma
is by an order of magnitude or more higher than that of a confined
hadronic medium, and so the energy loss of a fast passing color charge 
is expected to be correspondingly higher as well. Let us consider
this in more detail.

\medskip

An electric charge, passing through matter containing other bound or
unbound charges, loses energy by scattering. For charges of low incident
energy $E$, the energy loss is largely due to ionization of the target
matter. For sufficiently high energies, the incident charge scatters
directly on the charges in matter and as a result radiates photons of
average energy $\omega \sim E$. Per unit length of matter, the
`radiative' energy loss due to successive scatterings,
\be
-{dE \over dz} \sim E \label{9}
\ee
is thus proportional to the incident energy.

\medskip

This probabilistic picture of independent successive scatterings breaks
down at very high incident energies \cite{LPM}. The squared amplitude
for $n$ scatterings now no longer factorizes into $n$ interactions;
instead, there is destructive interference, which for a regular medium
(crystal) leads to a complete cancellation of all photon emission except
for the first and last of the $n$ photons. This Landau-Pomeranchuk-Migdal 
(LPM) effect greatly reduces the radiative energy loss.

\medskip

\begin{figure}[htb]
\vspace*{-0mm}
\centerline{\psfig{file=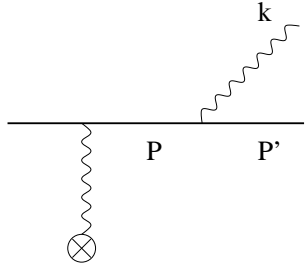,height= 40mm,angle=-90}}
\caption{Gluon emission after scattering.}
\label{4_4}
\end{figure}

The physics of the LPM effect is clearly relevant in calculating the
energy loss for fast color charges in QCD media. These media are not
regular crystals, so that the cancellation becomes only partial. Let us
consider the effect here in a heuristic fashion; for details of the
actual calculations, see \cite{Baier,Zak}. The time $t_c$ needed for the
emission of a gluon after the scattering of a quark (see Fig.
\ref{4_4}) is given by
\noindent
\be
t_c = {1 \over \sqrt{P^2}}{E \over \sqrt{P^2}} =
{E\over 2P'k}, \label{9a}
\ee
in the rest system of the scattering center, where $P^2$ measures 
how far the intermediate quark state
is off-shell; on-shell quarks and gluons are assumed to be massless, and
$E/\sqrt{P^2}$ is the $\gamma$-factor between the lab frame and the
proper frame of the intermediate quark. For gluons with $k_L >> k_T$,
we thus get
\be
t_c \simeq {\omega \over k_T^2}. \label{9b}
\ee
If the passing color charge can interact with several scattering
centers during the formation time of a gluon, the
corresponding amplitudes interfere destructively, so
that in effect after the passage of $n$ centers over the
coherence length $z_c$, only one gluon is emitted, in contrast to
the emission of $n$ gluons in the incoherent regime. Nevertheless,
in both cases each scattering leads to a $k_T$-kick of the charge,
so that after a random walk past $n$ centers, $k_T^2 \sim n$. Hence
\be
k_T^2 \simeq \mu^2 {z_c\over \lambda}, \label{9c}
\ee
where $\lambda$ is the mean free path of the charge in the medium, so
that $z_c/ \lambda >1$ counts the number of scatterings. At
each scattering, the transverse kick received is measured by the mass
of the gluon exchanged between the charge and the scattering center, i.e., by
the screening mass $\mu$ of the medium. From Eq.\ \ref{9b} we have
\be
z_c \simeq {\omega \over k_T^2}, \label{9d}
\ee
so that the formation length in a medium characterized by $\mu$ and
$\lambda$ becomes
\be
z_c \simeq \sqrt{{\lambda \over \mu^2} \omega}. \label{9e}
\ee
For the validity of Eq.\ (\ref{9e}), the mean free path has to be larger
than the interaction range of the centers, i.e., $\lambda > \mu^{-1}$.

\medskip

The energy loss of the passing color charge is now determined by
the relative scales of the process. If $\lambda >z_c$, we have
incoherence, while for $\lambda < z_c$ there is coherent scattering with
destructive interference. In both cases, we have assumed that the
thickness $L$ of the medium is larger than all other scales. When
the coherence length reaches the size of the system, $z_c = L$,
effectively only one gluon can be emitted. This defines a critical
thickness $L_c(E)=(E \lambda / \mu^2)^{1/2}$ at fixed incident energy
$E$, or equivalently a critical $E_c=\mu^2 L^2 /\lambda$ for fixed
thickness $L$; for $L > L_c$, there is bulk LPM-behavior, below $L_c$
there are finite-size corrections.

\medskip

\begin{figure}[htb]
\centerline{\psfig{file=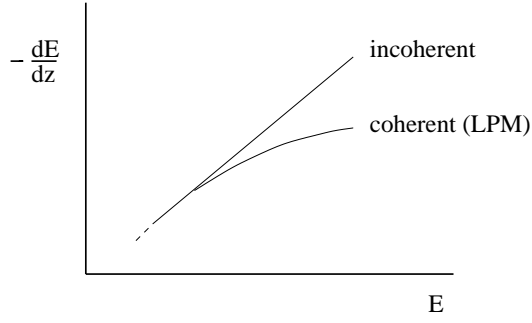,height=70mm,angle=-90}\hspace*{3mm}}
\caption{Energy loss in incoherent and coherent interactions.}
\label{4_5}
\end{figure}

\begin{figure}[htb]
\centerline{\psfig{file=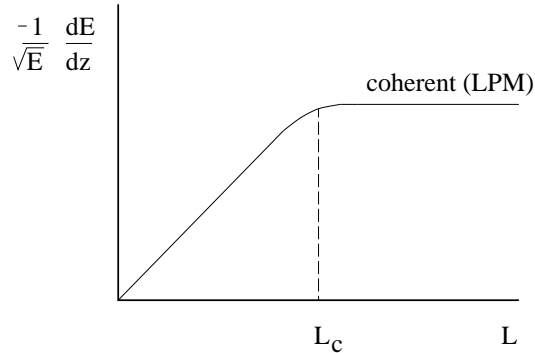,height=70mm,angle=-90}}
\caption{Energy loss in coherent interactions
as function of the thickness $L$ of the medium}
\label{4_6}
\end{figure}

We are thus left with three regimes for radiative energy loss. In case
of incoherence, $z_c < \mu^{-1}$, there is the classical radiative loss
\be
-{dE \over dz} \simeq {3 \alpha_s \over \pi} {E \over \lambda},
\label{9f}
\ee
where $\alpha_s$ is the strong coupling.
In the coherent region, $\lambda >z_c$, the energy loss is given by the
LPM bulk expression when $L > L_c$ \cite{Baier},
\be
-{dE \over dz} \simeq {3 \alpha_s \over \pi} \sqrt{{\mu^2 E \over \lambda}}.
\label{10}
\ee
The resulting reduction in the radiative energy loss $dE/dz$ is
illustrated in Fig.\ \ref{4_5}. Note that in earlier estimates
the energy loss due to interactions of the gluon cloud accompanying the
passing color charge had been neglected \cite{G-W}; this led to a
considerably smaller energy loss, proportional to $\ln E$ instead of
$\sqrt E$. Finally, in a medium of thickness $L < L_c$, there is
less scattering and hence still less energy loss. Eq.\ (\ref{10}) can be
rewritten as
\be
-{dE \over dz} \simeq {3 \alpha_s  \over \pi}{\mu^2 \over \lambda} L_c(E),
\label{11}
\ee
and for $L<L_c$, this leads to
\be
-{dE \over dz} \simeq {3 \alpha_s \over \pi} {\mu^2 \over \lambda} L
\label{12}
\ee
as the energy loss in finite size media with $L \leq L_c$.
The resulting variation of the radiative energy loss with the
thickness of the medium is shown in Fig.\ \ref{4_6}, with
saturated (i.e., bulk) LPM behavior setting in for $L\geq L_c$.

\medskip

Eq.\ (\ref{12}) has been used to compare the energy loss in a deconfined
medium of temperature $T =0.25$ GeV to that in cold nuclear matter of
standard density \cite{Schiff}. For the traversal of a medium of 10 fm
thickness, estimates give for the total energy loss
\be
\Delta E = \int_{0~\rm fm}^{10~\rm fm} dz {dE \over dz} \label{12a}
\ee
in a quark-gluon plasma
\be
-\Delta E_{qgp}  \simeq 30~~{\rm GeV},
\label{13}
\ee
corresponding to an average loss of 3 GeV/fm. In contrast, cold
nuclear matter leads to
\be
-\Delta E_{cnm} \simeq 2~~{\rm GeV}
\label{14}
\ee
and hence an average loss of 0.2 GeV/fm. A deconfined medium thus leads
to a very much higher rate of jet quenching than confined hadronic
matter, as had in fact been suggested quite some time ago \cite{Bj}.

\subsection{Initial State Considerations}

In using quarkonia and jets as tools, we have so far again considered a 
simplified situtation, in which we test a given medium with distinct
external probes. In heavy ion collisions, we have to create the probe
in the same collision in which we create the medium.
Quarkonia and jets (as well as open charm/beauty and very energetic dileptons 
and photons) constitute so-called ``hard probes'', whose production 
occurs at the very early stages of the collision, before the medium is
formed; they are therefore present when it appears. Moreover,
their production involves large energy/momentum scales and can 
be calculated by perturbative QCD techniques; the results can be 
tested in $pp/pA$ collisions, so that behavior and strength of such 
outside ``beams'' or ``color charges'' are in principle under control.

\medskip

On the other hand, such calculations based on hard partonic interactions
assume 
\begin{itemize}
\vspace*{-0.2cm}
\item{that the parton distributions function in nuclei are known, and}
\vspace*{-0.2cm}
\item{that the parton model itself is applicable for nuclear collisions.}
\vspace*{-0.2cm}
\end{itemize}
Both assumptions cannot be universally valid. The parton distribution
functions are modified in nuclei because of the presence of other parton
sources (shadowing, anti-shadowing), and these effects are to a considerable 
extent of non-perturbative nature. Moreover, the number of partons, and 
hence their density in the transverse plane, increase with collision
energy. Partons with an intrinsic transverse momentum have an intrinsic
size in the transverse plane, and so increasing their density will eventually
lead to parton saturation. This is another instance of the percolation
process discussed above. At this point of ``parton saturation'', any model 
of independent partonic interactions breaks down, and we have a new
medium. The study of such saturation effects has in recent years
attracted much attention (``color glass condensate'') 
\cite{CGC,Jai} and will be dealt 
with in other chapters of this handbook. 

\medskip

As already mentioned, another crucial aspect for the formation of a
quark-gluon plasma in high energy nuclear collisions is the question
of how a locally equilibrated medium can be formed from a non-equilibrium
initial state. This question arises for a partonic initial state 
(parton thermalization) as well for the possible transition of a 
primary saturated medium to a quark-gluon plasma, with a possible 
futher intermediate state (``glasma'') \cite{Jai,glasma}. Here again we 
refer to subsequent chapters.
  
\section{Summary}

We have shown that strong interaction thermodynamics results in a well-defined
transition from hadronic matter to a plasma of deconfined quarks and gluons.
For vanishing baryon number density, the transition provides both
deconfinement and chiral symmetry restoration at $T_c \simeq 160 - 190$ MeV.
At this point, the energy density increases by an order of magnitude through
the latent heat of deconfinement.

\medskip

The behavior of strongly interacting matter for increasing baryon density
is presently at the focus of much attention; both the change of the 
transition nature with $\mu$ and the origin for the expected changes
have to be clarified further. 

\medskip

The properties of the new medium above $T_c$, the quark-gluon plasma, can 
be studied through hard probes (quarkonium dissociation, jet quenching) 
and electromagnetic radiation (photons and dileptons). Information about 
transition aspects is provided by light hadron radiation; in particular,
experimental species abundances show a universal hadronization temperature 
in accord with that found in non-perturbative QCD studies.

\end{document}